\documentclass[aps,prc,twocolumn,letterpaper,superscriptaddress,nofootinbib,showpacs,floatfix,10pt]{revtex4-1}
\usepackage[utf8]{inputenc}
\usepackage{graphicx}
\usepackage{comment}
\usepackage{setspace}
\usepackage{amsmath}
\usepackage{amssymb}
\usepackage[textwidth=16cm,textheight=22cm]{geometry}
\usepackage{color}
\usepackage{indentfirst}
\usepackage{xspace}
\usepackage{hyperref}
\usepackage{verbatim}
\usepackage{epstopdf}
\usepackage{graphicx}
\usepackage{longtable}
\usepackage{lineno}
\usepackage{subfigure}
\usepackage{parskip}
\usepackage{xcolor}

\newcommand{\pt}   {\ensuremath{p_\mathrm{T}}\xspace}

\newcommand{\nch}   {\ensuremath{N_\text{ch}}\xspace}
\newcommand{\pythia} {\textsc{pythia8}\xspace}
\newcommand{\epos} {\textsc{epos-lhc}\xspace}

\begin{document}


\title{
System size and event shape dependence of particle-identified balance functions in proton-proton collisions at $\sqrt{s}=13$ TeV using PYTHIA 8 and EPOS models
}

\author{Subash Chandra Behera}
\email{subash.chandra.behera@cern.ch}
\affiliation{
INFN--Sezione di Roma, Piazzale Aldo Moro, 2 - 00185 Roma RM, Italy}
\author{Arvind Khuntia}
\email{arvind.khuntia@cern.ch}

\affiliation{INFN--Sezione di Bologna, via Irnerio 46, 40126 Bologna BO, Italy}

\begin{abstract}
We investigate charge balance functions for pion, kaon, and proton pairs in proton-proton (pp) collisions at $\sqrt{s}=13$~TeV using Monte Carlo models, \pythia and \epos, with transverse spherocity to classify event topology and charged-particle multiplicity to select system size. Simulations with \pythia and \epos reveal that balance-function widths in rapidity and azimuthal angle depend on multiplicity and event shape. In \pythia, widths decrease monotonically with multiplicity, consistent with local charge conservation in a fragmentation-dominated scenario. In contrast, the \epos model, especially when using the core corona implementation, exhibits a more intricate response, where the combined effects of hydrodynamic radial flow and longitudinal diffusion result in narrower azimuthal correlations and broader rapidity correlations. These features are characteristic signatures of collective dynamics, similar to those observed in heavy-ion collisions. Events with low spherocity, which are jet-like in nature, exhibit significantly narrower balance function widths compared to isotropic events with high spherocity, illustrating that event-shape selection provides clear sensitivity to the underlying dynamics of particle production in pp collisions. The species dependence and event-shape sensitivity of the balance-function widths provide information about the hadronization dynamics and collectivity in small systems. These results demonstrate that multidimensional, particle species dependent balance function measurements can disentangle the underlying mechanisms of charge correlations and medium-like behavior in high-multiplicity pp collisions.
\end{abstract}

\keywords{Quark-Gluon Plasma, spherocity, small system, collectivity, balance function}

\maketitle

\section{Introduction}
An exotic state of matter known as quark-gluon plasma (QGP) is believed to form under extreme conditions created in ultra-relativistic heavy-ion collisions at the Large Hadron Collider and the Relativistic Heavy Ion Collider~\cite{QGPmedium1, starqgp, qgpmed2, cmswhite}. Traditionally, signals of QGP were anticipated only in large systems such as nucleus–nucleus collisions. Recent experimental results, however, have revealed unexpected signs of collective behavior in smaller systems like proton–proton (pp) and proton–nucleus (p–A) collisions, a feature previously thought to be exclusive to medium formation~\cite{cmsppridge, cmsppflow, Baty:2021ugw, CMS:2016fnw, ATLAS:2023bmp, ATLAS:2019wzn, ATLAS:2016yzd, ATLAS:2015hzw, ATLAS:2014qaj, ATLAS:2012cix, ALICE:2012eyl, STAR:2023wmd, STAR:2022pfn, PHENIX:2021ubk, Mace:2018vwq}. A central challenge in high-energy physics is to determine whether the observed collective-like behavior in small systems arises from true macroscopic, medium-like phenomena such as hydrodynamic flow or can be explained by purely microscopic QCD mechanisms like string fragmentation, color reconnection, and multiparton interactions. The balance function ($B$) is particularly sensitive to this distinction as its shape and width encode information about the timing, spatial separation, and collective motion of balancing charges~\cite{Basu:2020ldt, scottpratt, bfscott, Pruneau:2019baa, ALICE:pruneau, Manea:2024qgd, alicebfpbpbpb,alicep2r2pp, alicethree, BassScott, cmsbf}. 

Hydrodynamic evolution, as implemented in models like \epos~\cite{epos_ref, eposLHC}, can induce collective flow and diffusion effects through the formation and expansion of a dense core medium, which enables the simulation of medium-like phenomena even in small collision systems. In contrast, microscopic models such as \pythia~\cite{pythiaref, pythia_ref}  provide a baseline scenario for charge correlations that arise without an explicit final-state medium. In \pythia, the production and fragmentation of color strings govern hadronization, with color reconnection algorithms allowing for the rearrangement of color connections between partons to minimize string length, while the spatial overlap of strings in high-multiplicity events can further influence particle production. Together, these microscopic effects in \textsc{pythia} serve as an alternative to collective behavior, offering a contrasting perspective to the medium-induced dynamics modeled in \epos.

The balance function is a statistical tool designed to study the correlations by measuring the conditional probability of observing a particle of opposite charge relative to a reference particle, as a function of rapidity ($y$) or azimuthal angle ($\phi$) differences~\cite{cmsbf, hydjetbf, scottpratt, ALICE:pruneau, Manea:2024qgd, aliceoldpbpb, STARBF1, STARBF2, Pratt:theo, scottpratt, BassScott}. The $B$ was initially introduced to explore the hadronization time and the sequential nature of quark production in the QGP. This framework proposes that quark–antiquark pairs are created in two distinct phases~\cite{Pratt:2012dz, scottpratt, cmsbf}. During the collision, quark–antiquark pairs are produced and subsequently hadronize into mesons and baryons. Due to the charge conservation, every quark produced must be accompanied by a corresponding antiquark, and these balancing charges are usually found close together in rapidity space. Theoretical models suggest that quark-antiquark pairs are not all created simultaneously. Instead, production happens in two main phases. The first wave occurs shortly after the collision begins, as the initial system thermalizes to QGP. The second phase takes place later, during hadronization, when the system cools and quarks combine to form hadrons. Entropy conservation at this stage implies an approximately constant number of effective quasiparticles, and because each hadron contains at least two quarks, most quark production occurs during this phase.

Quark–antiquark pairs that are produced early in the collision, and have more time to separate spatially before hadronization, lead to broader correlations in the balance function. In contrast, if the quark-antiquark pairs are produced later in the system's evolution, they tend to remain closer together. At the same time, the development of strong radial flow can also influence particle correlations by pushing balancing partners into similar directions in momentum space, effectively narrowing the balance function. This collective expansion tends to align balancing particle pairs more closely in momentum space, reducing their separation in rapidity. Notably, the balance function width exhibits an inverse relationship with the transverse mass, $m_\mathrm{T} = \sqrt{m^{2} + p_\mathrm{T}^{2}}$, where higher $m_\mathrm{T}$ particles are associated with narrower correlations~\cite{icpaqgp, cmsbf}.

Therefore, a comprehensive study of particle species dependent balance functions is essential for disentangling the microscopic mechanisms of charge production and transport in small collision systems, where the dynamics are constrained by local charge conservation~\cite{scottpratt,bfscott, alicethree, aliceoldpbpb, icpaqgp}. By measuring balance functions separately for pions, kaons, and protons, we gain unique access to the distinct production and transport dynamics of light, strange, and baryonic charges, respectively. Pions, as the lightest and most copiously produced hadrons, predominantly reflect late-stage dynamics and global charge conservation. Kaons, carrying strangeness, are sensitive probes of the medium’s strangeness content and the role of associated production mechanisms. Protons and antiprotons, meanwhile, are particularly useful for studying the transport of conserved baryon number across rapidity and the role of annihilation processes in the hadronic phase~\cite{scottpratt, ALICE:pruneau, scottpratt}. Comparative analysis of particle-identified (PID) balance functions allows us to distinguish between early- and late-stage charge creation, test the degree of collectivity and flow-like effects for different particle species, and place stringent constraints on theoretical models of hadronization and medium evolution. This differential approach is particularly important for building a comprehensive picture of how collective phenomena and charge correlations manifest across the full spectrum of hadron species in small collision systems.  To deepen our understanding of the underlying dynamics, it is essential to investigate balance functions as a function of charged-particle multiplicity ($\langle \nch \rangle$) and event topology, which is quantified using transverse spherocity ($S_{0}$)~\cite{ALICE_SP1, arvind_sp}. Multiplicity reflects the overall activity and energy density in the collision, offering insight into how collective phenomena may emerge as the system size increases. Transverse spherocity, as an event-shape variable, allows us to distinguish between jet-like events that are dominated by hard scatterings and isotropic events that are more sensitive to soft or collective processes. Examining balance functions in this multidimensional framework helps to disentangle the relative roles of collective flow, multi-parton interactions, and jet fragmentation in shaping charge correlations in small systems.

In this paper, the balance functions are studied for different charged species, $\pi$, K and $\it{p}$, as a function of transverse spherocity using a large rapidity acceptance coverage of $|y| \leq 2.4$. The paper is organized as follows. Section~\ref{anaproedure} discusses the analysis technique to construct the correlation and balance functions. Section~\ref{modeldes} demonstrates the model calculations in \textsc{pythia8} and \textsc{epos}. Section~\ref{results} presents the results of the balance functions of $\pi$, K and $\it{p}$ in relative rapidity ($\Delta y$) and relative azimuthal angle ($\Delta\phi$), and their width in different $\langle \nch \rangle$ and $S_{0}$ classes~\cite{mult_def}. Section~\ref{summary} presents the summary of this work.

\section{Analysis Methodology}
\label{anaproedure}

The conservation of electric charge manifests through the correlated production of oppositely charged particle pairs. The charge balance function provides a differential measure of these correlations in phase space, offering sensitivity to charge creation and transport dynamics. Typically expressed as a function of $\Delta y$ and $\Delta\phi$, and quantifies the correlation strength between oppositely charged particles. Mathematically, this can be written as
\begin{equation} \label{eqn_balfun}
B^{\alpha \beta}=\frac{1}{2}[C_{2}^{\alpha^{+} \beta^{-}} + C_{2}^{\alpha^{-} \beta^{+}} - C_{2}^{\alpha^{-} \beta^{-}} - C_{2}^{\alpha^{+} \beta^{+}}],
\end{equation}

where \( \alpha \) and \( \beta \) denote the hadron species under consideration, and the superscripts \( + \) and \( - \) represent the particle charges. Each term \( C_2^{\alpha^q \beta^{q'}} \) corresponds to the two-particle correlation function for a specific charge combination. The first two terms represent correlations between unlike-sign pairs, which are expected to capture the primary effects of local charge conservation. The subtraction of like-sign pair correlations removes charge-independent correlations, enhancing the sensitivity to genuine balancing contributions. The balance functions for different combinations of identified hadrons (e.g., \( B^{\pi\pi} \), \( B^{KK} \), \( B^{pp} \)), can access the dynamics of light, strange, and baryonic charge conservation separately. For instance, \( B^{KK} \) probes strangeness correlations and is sensitive to the production and hadronization of strange quarks, while \( B^{pp} \) captures features associated with baryon number conservation and transport.

These correlation functions are derived from the ratio of signal and mixed-event distributions, which are calculated in $0.2 < \pt < 2.0$ GeV and a large acceptance coverage, $|y| < 2.4$, to probe the bulk particle production following standard methods in previous analyses~\cite{Baty:2021ugw, CMS:2023iam, cmsbf, cmsppridge, cmspbpbflow, hin18008, CMSPP}. The signal distribution, denoted as \( S(\Delta y, \Delta \phi) \), is constructed by pairing particles within the same event to capture genuine physical correlations. It is defined as:
\begin{equation} \label{eqn_sig}
S(\Delta y, \Delta \phi) = \frac{1}{N_\text{trig}} \frac{d^{2}N^\text{same}}{d\Delta y \, d\Delta \phi},
\end{equation}
where \( N_\text{trig} \) represents the number of trigger particles in the selected \( p_{\mathrm{T}} \) interval, $0.2 < \pt < 2.0$ GeV and $|y| \leq 2.4$. \( N^\text{same} \) is the number of particle pairs binned in $\Delta y$ and $\Delta\phi$. A conventional event-mixing technique is used to generate the mixed-event distribution, $M(\Delta y, \Delta\phi)$.  In this approach, trigger particles from a given event are combined with associated particles chosen from a set of ten randomly selected events within the same multiplicity class. The mixed event distribution is defined as \begin{equation} \label{eqn_mix}
M(\Delta y,\Delta \phi) = \frac{1}{N_\text{trig}}\frac{d^{2}N^\text{mix}}{{d\Delta y} \ {d\Delta \phi}},
\end{equation}
where the number of mixed event pairs for a given $\Delta y$ and $\Delta\phi$ bin is denoted by $N_\text{mix}$. The mixed event distribution corrects the correlation functions due to the finite detector acceptance. The angular correlation functions in two dimensions are defined as 
\begin{equation} \label{eqn_2pc}
\frac{1}{N_\text{trig}}\frac{d^{2}N^\text{pair}}{{d\Delta y} \ {d\Delta \phi}} = M(0,0) \frac{S(\Delta y, \Delta\phi)}{M(\Delta y, \Delta\phi)}.
\end{equation}
The effect of pair acceptance is largely corrected by applying the ratio  $M(0,0)/M(\Delta y, \Delta\phi)$. $M(0,0)$ represents the mixed-event pair yield for particles emitted in almost the same direction, and corresponds to the region of largest pair-acceptance~\cite{cmspbpbflow, cmsbf, cmsppridge}.
Like-sign and unlike-sign correlations manifest different physics phenomena such as collective flow, a contribution due to minijet component, resonance decay, Coulomb attraction and Coulomb repulsions~\cite{cmspbpbflow, cmsbf}, quantum-statistical effects such as Bose–Einstein type of correlations for identical pairs~\cite{Pratt_ref1, Brown_ref2}, and charge conservation effect. 

\begin{figure}[htbp]
  \centering
  \includegraphics[width=0.49\textwidth]{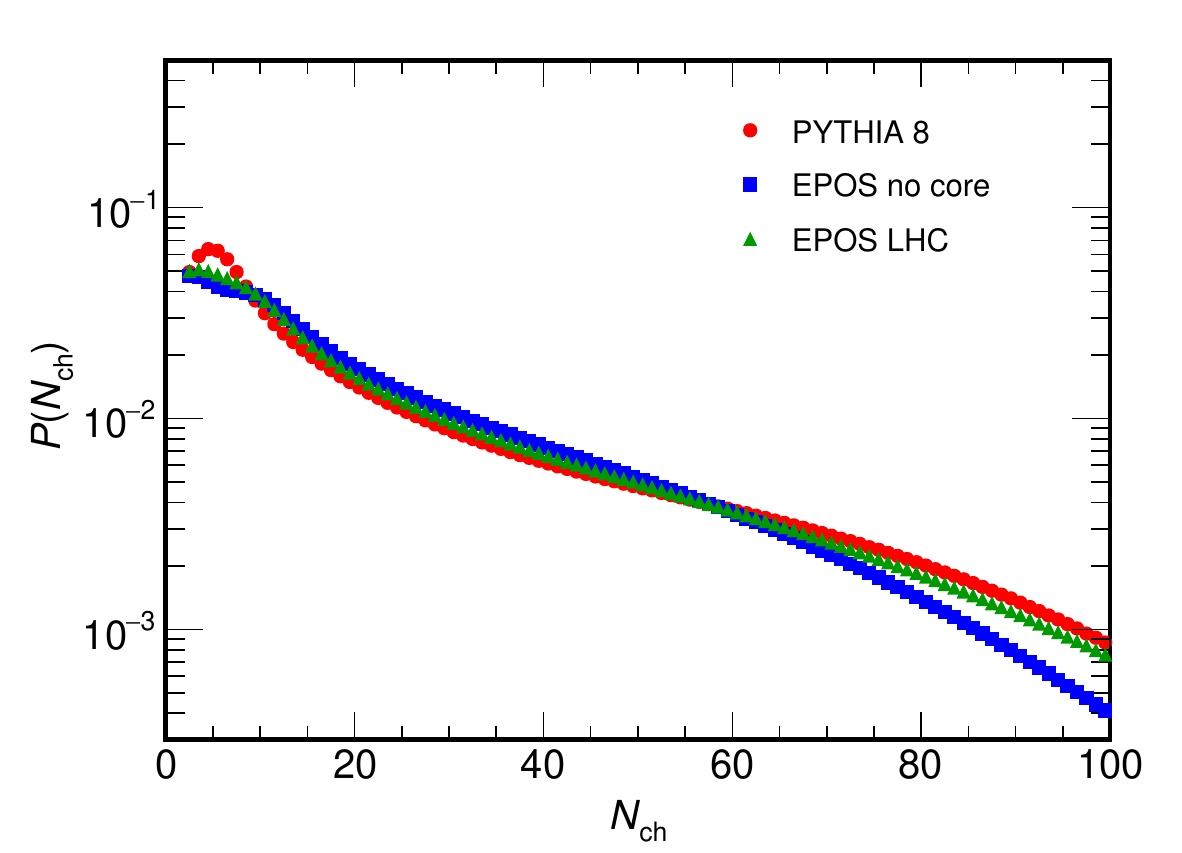}
  \caption{Charged particle multiplicity distributions for \pythia and \textsc{epos} event generators in pp collisions at $\sqrt{s} = 13$ TeV.}
  \label{fig:nch_eta}
\end{figure}
\begin{table}[ht]
\centering
\caption{Percentage of events within the selected multiplicity classes for \pythia, \textsc{epos} no core and \epos.}
\vspace{0.3cm}
\begin{tabular}{c|c|c|c}
\hline
\nch  & \pythia & \textsc{epos} no core & \epos \\
\hline
$0-20$   & 63.37  & 60.39  & 61.64 \\
$20-40$  & 18.57 & 22.66 & 20.53 \\
$40-60$  & 9.59 & 10.58 & 10.05 \\
$60-80$  & 5.67 & 4.72 & 5.32 \\
$80-100$ & 2.78 & 1.62 & 2.44 \\
\hline
\end{tabular}
\label{tab:nch}
\end{table}

In this paper, balance functions are studied using the event topology variable, transverse spherocity, which classifies events based on back-to-back jet topologies to that of an isotropic event using the particle distributions of final-state, which arises from hadronic and nuclear collisions. Mathematically, spherocity can be written as, 
\begin{equation}
    S_{0} = \frac{\pi^2}{4} \min_{\hat{n}} \left( \frac{\sum_i |\vec{p}_{\rm T,i} \times \hat{n}|}{\sum_i \vec{p}_{\rm T,i}} \right)^2.
\end{equation}
The unit vector $\hat{n}$ is used in this case to minimize the ratio $S_{0}$ inside the brackets. $\frac{\pi^2}{4}$ is a scaling factor which ensures that the $S_{0}$ estimator falls between 0 and 1. At mid-rapidity, we have considered at least five charged particles in $|y|<1$ to have a meaningful spherocity definition~\cite{arvind_sp}. 
\begin{figure*}[!ht]  
    \centering
    \subfigure[]{
        \includegraphics[width=0.45\textwidth]{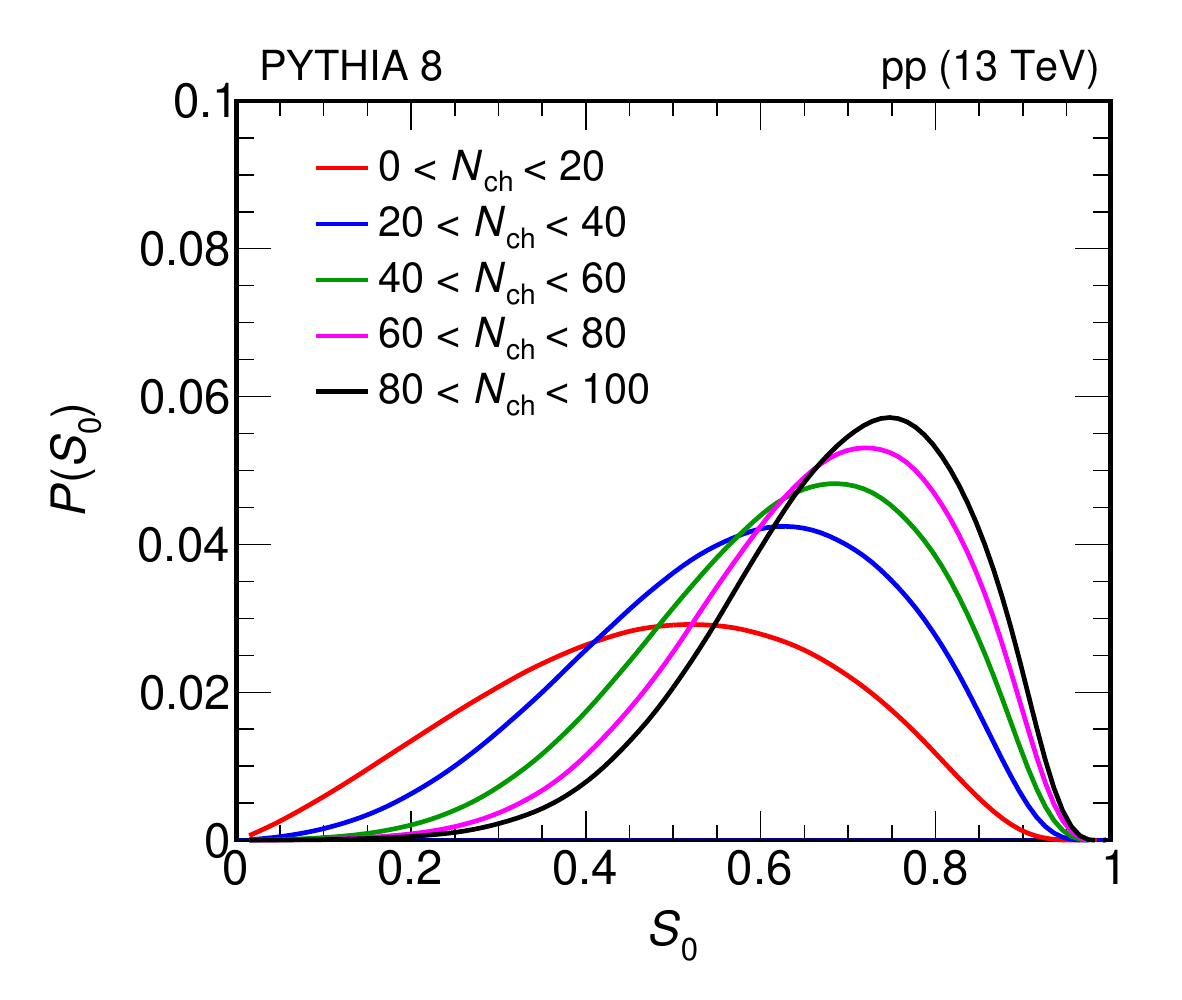}
    }
    \subfigure[]{
        \includegraphics[width=0.45\textwidth]{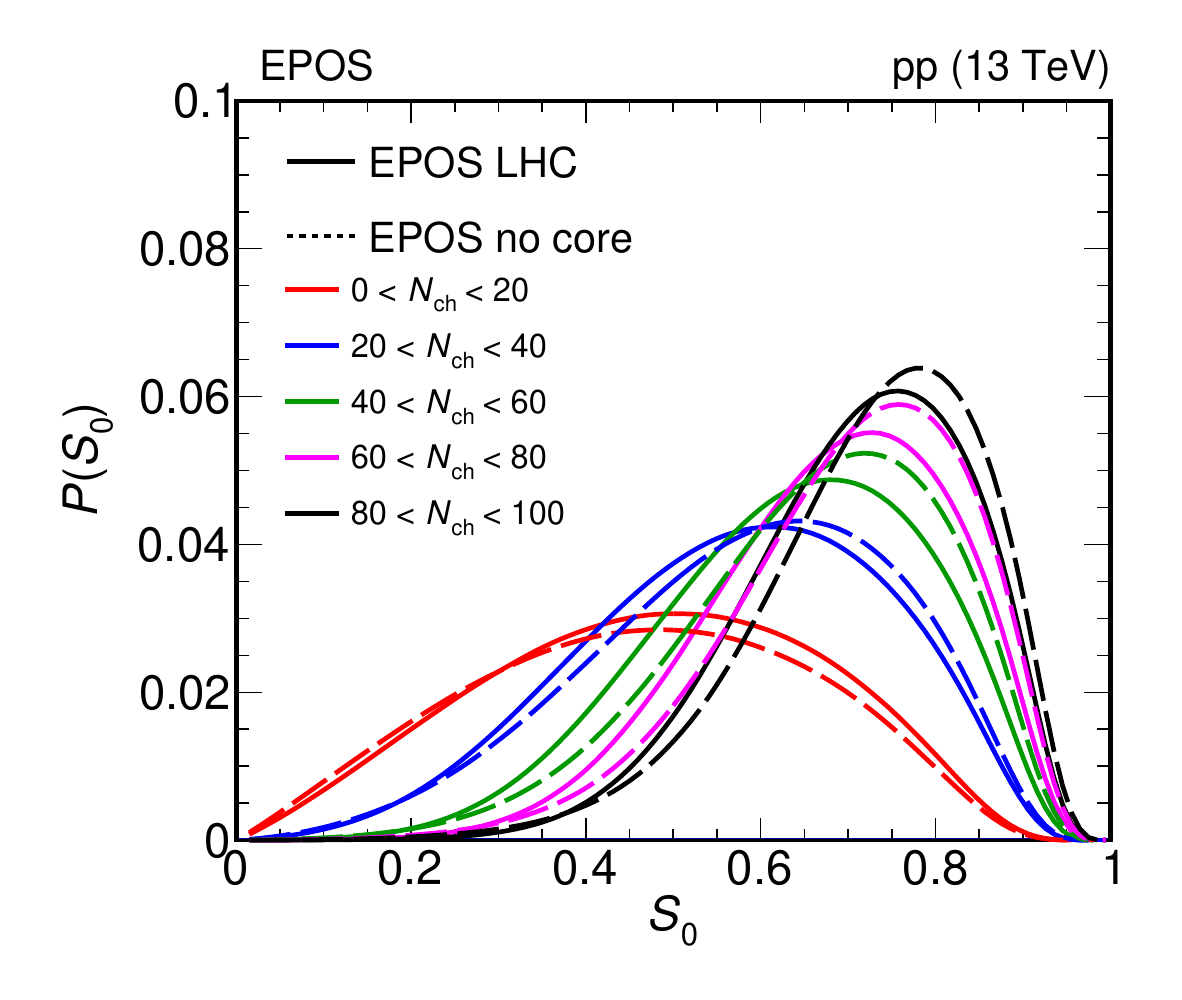}
    }
 \caption{The transverse spherocity distributions~\cite{arvind_sp, ALICE_SP1, ALICE_SP2} for different
multiplicity classes~\cite{mult_def} in pp collisions at $\sqrt{s} = 13$ TeV using the
\pythia (left) and \textsc{epos} (right) event generators.}
    \label{fig:sp_epos_pythia}
\end{figure*}

To quantitatively measure the width of charge-dependent correlations in a given variable \( \Omega \), the width is determined using the root-mean-square (RMS) method, assuming that the balance function is centered around zero \cite{alicebfpbpbpb, alicep2r2pp}:
\begin{equation}
\sigma = \left[ \frac{\sum_i O(\Omega_i)\, \Omega_i^2}{\sum_i O(\Omega_i)} \right]^{\frac{1}{2}},
\label{eq:bf_width}
\end{equation}
where \( O(\Omega_i) \) represents the value of the balance function at the bin centered at \( \Omega_i \), and the summation runs over all bins of the measured distribution. Here, \( \Omega\) denotes the quantity ($\Delta y$ or $\Delta \phi$) along which the balance function width is being evaluated. The width of the $B$ is estimated in $|\Delta y| \leq 1.5$, and for the $|\Delta\phi| \leq 1.5$ range from the one-dimensional projection of $B(\Delta y, \Delta \phi)$.

\begin{figure*}[!htb]  
    \centering
    \subfigure[]{
        \includegraphics[width=1.0\textwidth]{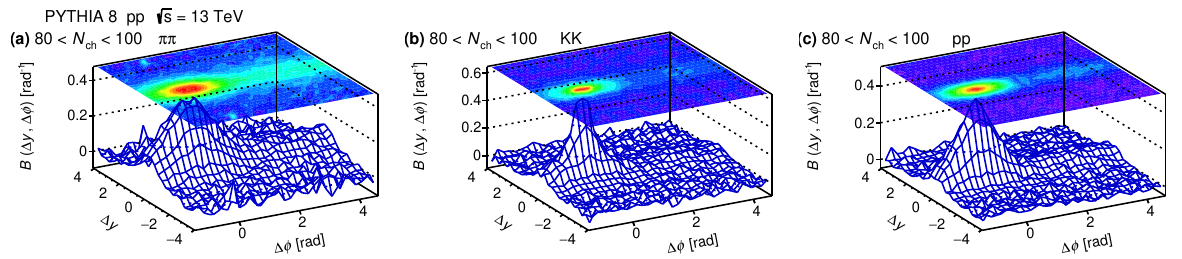}
    }
    \subfigure[]{
        \includegraphics[width=1.0\textwidth]{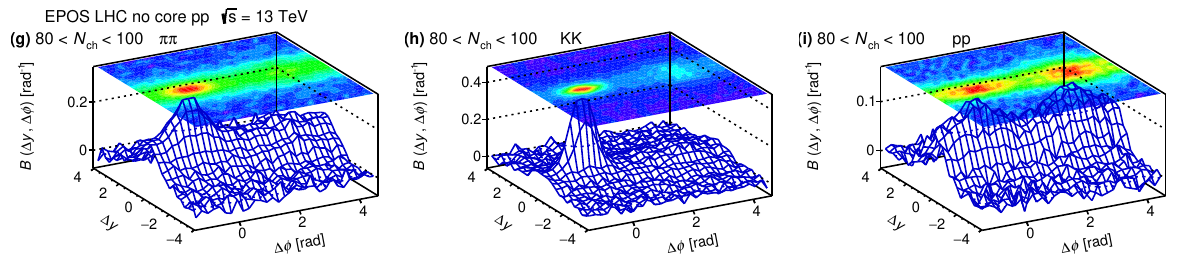}
    }
    \subfigure[]{
        \includegraphics[width=1.0\textwidth]{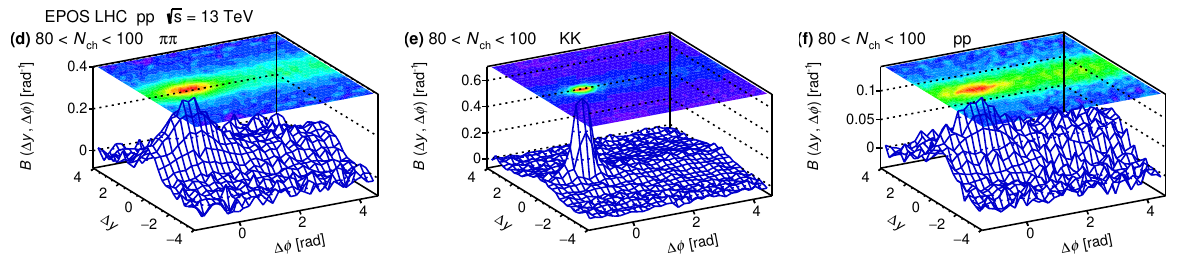}
    }
 \caption{Two-dimensional balance function from \pythia and \epos model simulation for $\pi, K$ and $p$ in pp collisions at $\sqrt{s} = $13 TeV. Plots are shown for the integrated spherocity value in high multiplicity collisions, $ 80 < N_{\mathrm{ch}} < 100$.}
    \label{fig:cbf_epos_pythia}
\end{figure*}

\section{Model description}
\label{modeldes}
In this measurement, we use two popular Monte Carlo (MC) event generators: \epos~\cite{eposLHC, epos_ref} and \pythia~\cite{pythia_ref, pythiaref}. A total of one billion pp events at $\sqrt{s} = 13$ TeV were simulated to study the multiplicity and spherocity dependence of balance functions for $\pi$, $K$ and $p$. \textsc{pythia8.306} version with the CP5 tune, which has been fine-tuned to describe a broad range of LHC observables, is employed in this study~\cite{CMS:2022awf}. In this configuration, soft QCD inelastic interactions are simulated to model minimum-bias pp collisions. It is based on the NNPDF3.1 parton distribution functions evaluated at next-to-next-to-leading order (NNLO)~\cite{CMS:2022awf}. The key mechanisms involved in the modeling of soft QCD particle production, such as multi-parton interactions (MPI) and color reconnection (CR), are included to replicate the complex, high-density partonic environment typically associated with high-multiplicity pp events~\cite{pythia_ref, pythiaref}. In \pythia, the multi-parton interaction framework models the occurrence of several parton-parton scatterings within a single proton-proton collision, which has a substantial impact on both the underlying event activity and the resulting particle production. The CR mechanism accounts for the rearrangement of color connections among partons before hadronization, effectively shortening color string lengths and leading to collective-like phenomena such as radial flow~\cite{pythia_ref, OrtizVelasquez:2013ofg}. These features are important for reproducing the observed characteristics of two-particle correlation structures, mainly in events with large particle multiplicities.

\epos, on the other hand, implements a hybrid modeling strategy that bridges microscopic parton-level processes and macroscopic collective dynamics. It starts with a parton-based initial state that evolves through multiple scatterings, followed by the classification of produced matter into two components: a dense core and a dilute corona. The core undergoes a collective expansion described by viscous hydrodynamics, while the corona hadronizes independently. This core-corona separation enables \epos to simulate radial flow and long-range correlations, even in small systems like pp collisions. Furthermore, the model incorporates conservation laws at every stage and includes non-linear effects such as parton saturation and string interactions. In this work, the \epos simulations are performed both with and without the hydrodynamic core enabled, allowing us to quantify the influence of collective expansion on the balance-function observables. By comparing the balance functions in these two models, we can explore the role of collective behavior, flavor conservation, and hadronization dynamics in shaping the charge correlations observed in data. \pythia provides a baseline without explicit collective flow, while \epos allows for an examination of flow-like effects and medium response in high-multiplicity pp events.

\begin{figure*}[!htb]  
    \centering
    \subfigure[]{
        \includegraphics[width=0.3\textwidth]{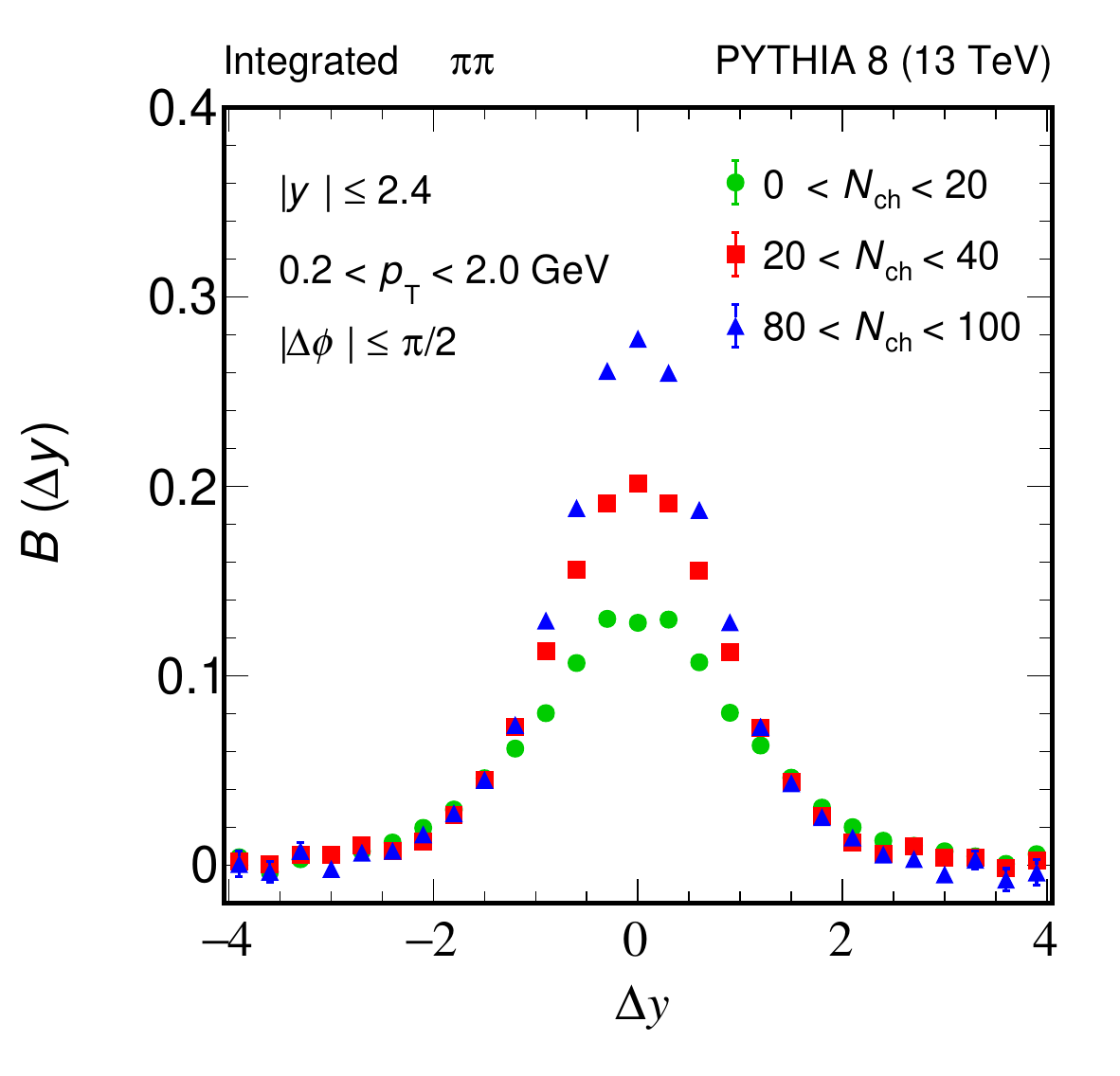}
    }
    \subfigure[]{
        \includegraphics[width=0.3\textwidth]{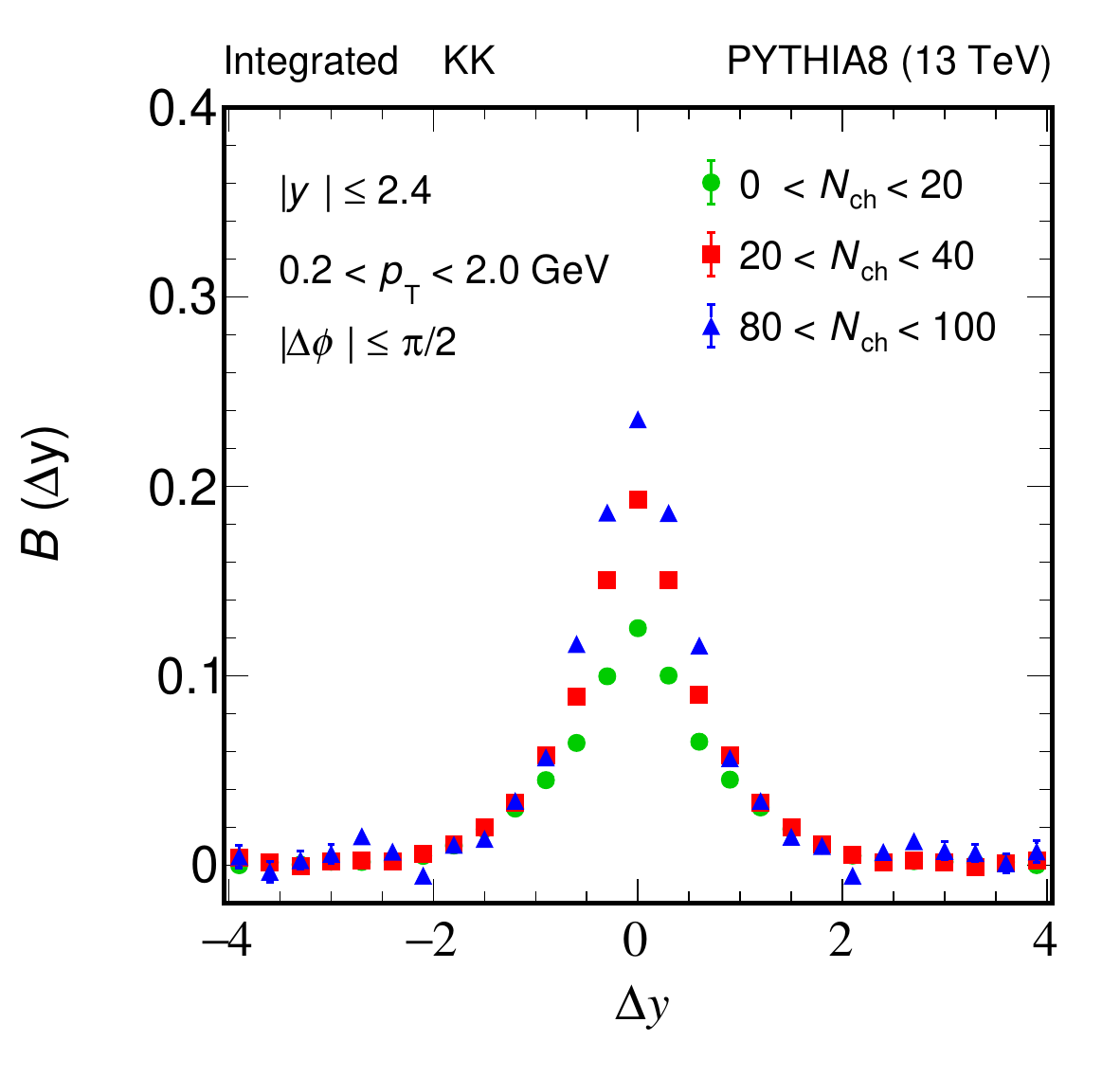}
    }
    \subfigure[]{
        \includegraphics[width=0.3\textwidth]{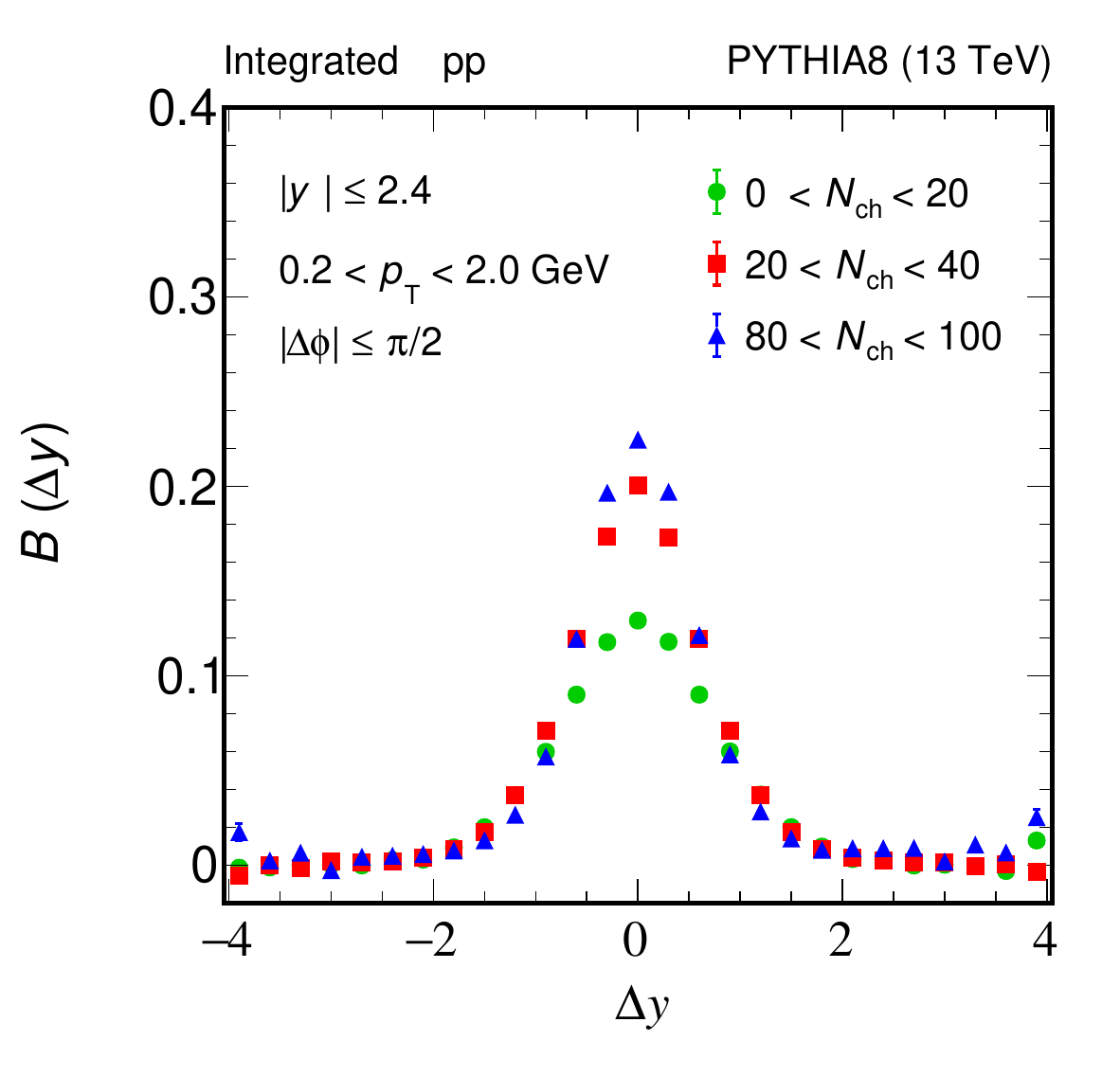} 
    }

    \subfigure[]{
        \includegraphics[width=0.3\textwidth]{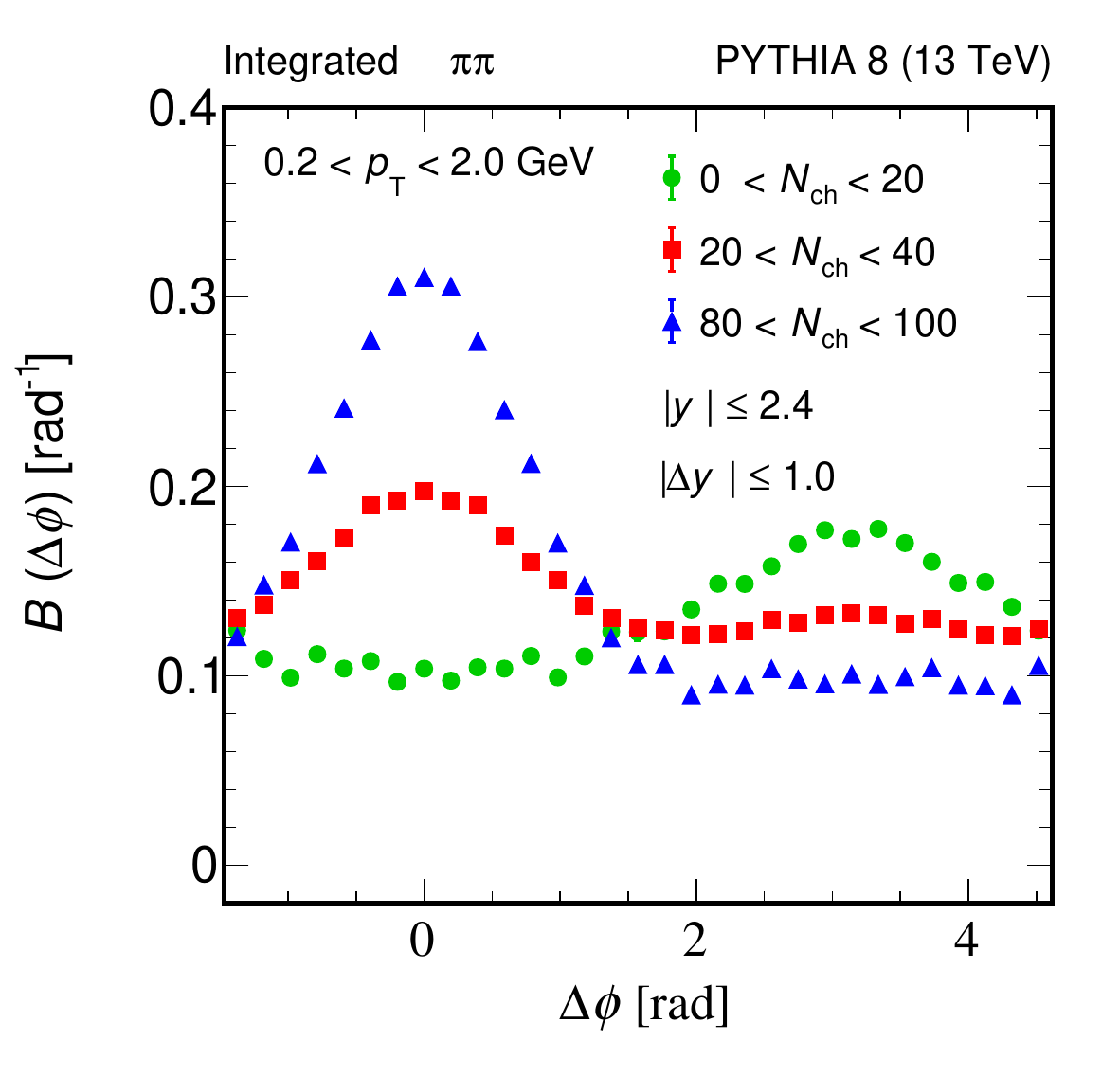}
    }
    \subfigure[]{
        \includegraphics[width=0.3\textwidth]{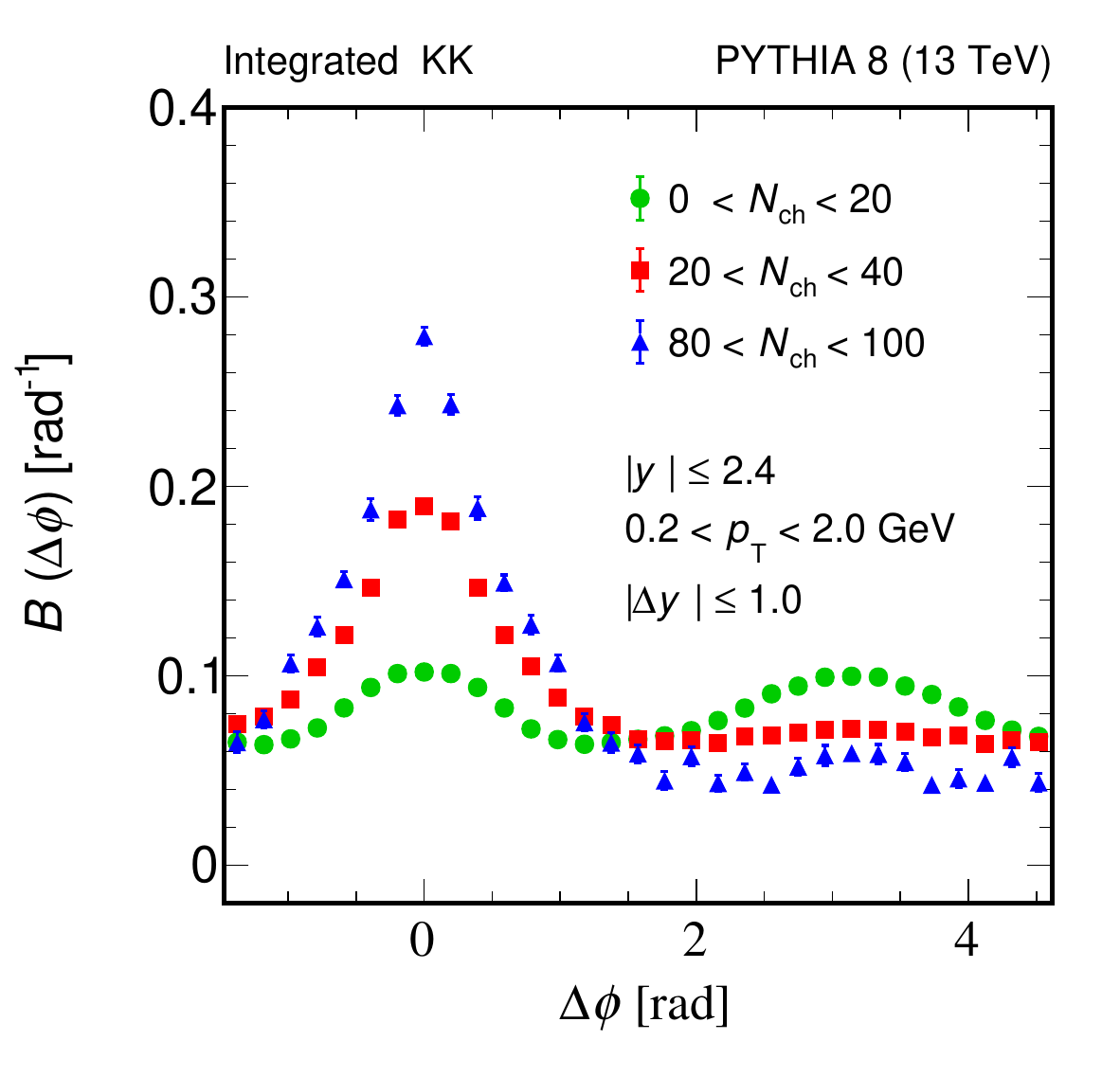} 
    }
    \subfigure[]{
        \includegraphics[width=0.3\textwidth]{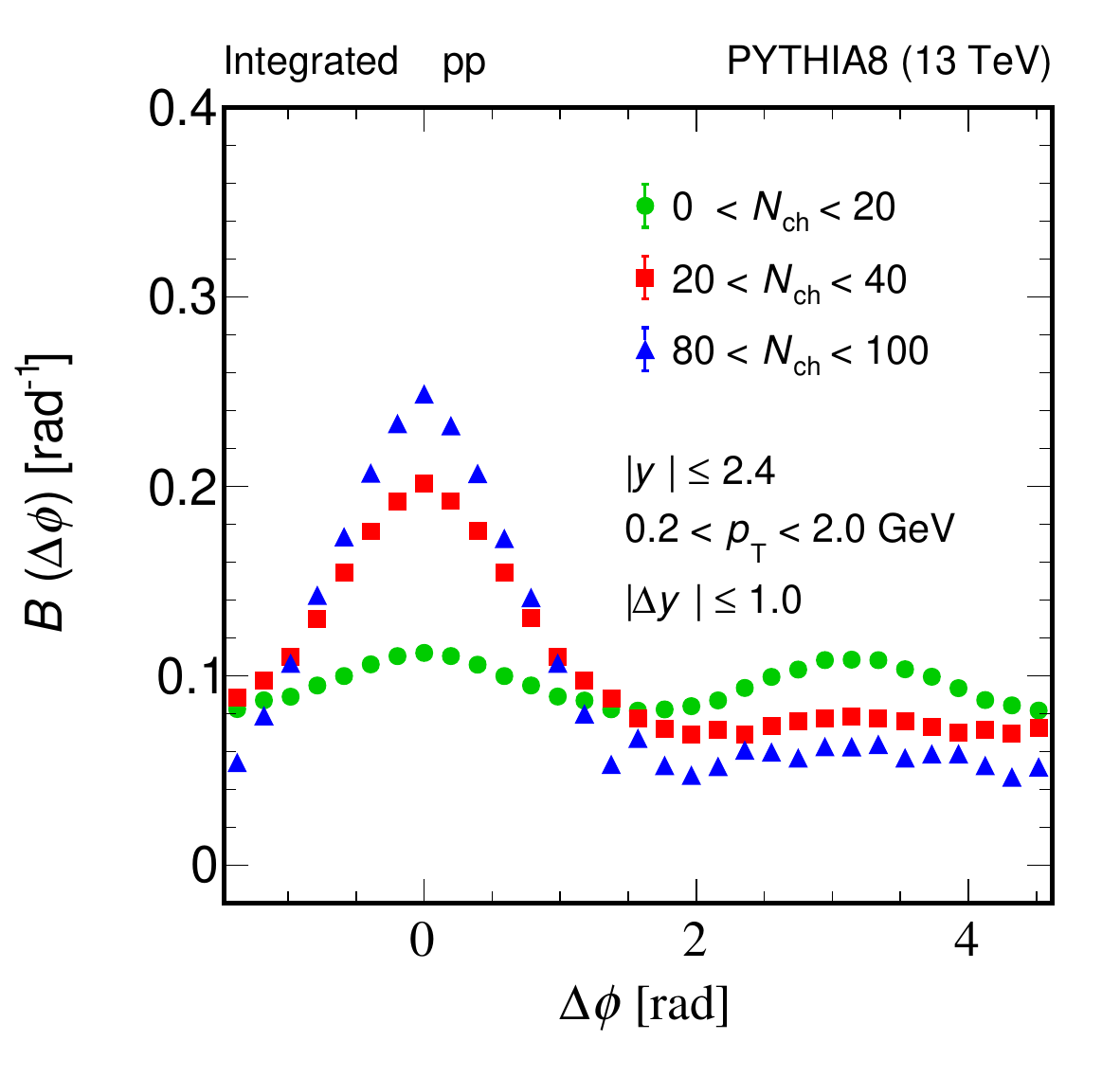} 
    }
 \caption{One-dimensional $B$ projections along $\Delta y$ and $\Delta\phi$ from \pythia model in pp collisions at $\sqrt{s} =$ 13 TeV. The left column is for pion, the middle column is for kaon, and the right panel is for proton, respectively. $\Delta y$ projections are take in $|\Delta\phi| \leq \pi/2$ and $\Delta\phi$ projections are taken in $|\Delta y| \leq 1.0$ range.}
    \label{fig:model_plots_1d}
\end{figure*}

\section{Result and discussion}
\label{results}

We first present the charged particle multiplicity distributions generated by \pythia, \textsc{epos} no core and \textsc{epos} studied in this work. For this purpose, the charged particle multiplicity $N_{\mathrm{ch}}$ is defined using primary charged particles within  $|y|<2.4$ and $p_{\mathrm{T}}>0.4~\mathrm{GeV}$, following the typical tracking acceptance of the CMS and ATLAS experiments. This definition establishes the multiplicity classes used throughout the analysis and ensures a consistent basis for comparing the different models. Figure~\ref{fig:nch_eta} shows the resulting 
$N_{\mathrm{ch}}$ distributions, while Table~\ref{tab:nch} summarizes the fraction of events populating each selected multiplicity interval. These provide the baseline for interpreting the subsequent balance-function measurements. Figure~\ref{fig:sp_epos_pythia} shows the transverse spherocity distributions for the 
selected charged-particle multiplicity classes in \pythia and in the two \textsc{epos} configurations. In all models, the $S_{0}$ spectra evolve systematically with increasing $N_{\mathrm{ch}}$, becoming progressively narrower and peaking at larger values of $S_{0}$, consistent with the increasing dominance of isotropic soft particle production at high multiplicity. These distributions provide the basis for selecting the jet-like (low $S_{0}$) and isotropic (high $S_{0}$) event classes used in the balance-function analysis.

\subsection{2D correlations}
Figure~\ref{fig:cbf_epos_pythia} shows the two-dimensional balance function distributions as functions of $\Delta y$ and $\Delta\phi$ for pions, kaons, and protons in the range $0.2 < p_{\mathrm{T}} < 2.0$ GeV and $|y| \leq 2.4$, comparing high-multiplicity events ($ 80 < N_{\mathrm{ch}} < 100$) from \pythia and \epos simulations. A clear distinction emerges between the two models: in \epos, particularly when the hydrodynamic core is included, the balance functions reveal a narrow and pronounced near-side peak centered at $(\Delta y, \Delta\phi) = (0,0)$, especially in the $\Delta\phi$ direction. This structure indicates strong short-range charge correlations and is attributed to collective effects such as radial flow, in addition to contributions from mini-jets, resonance decays, and local charge conservation. In contrast, \pythia produces broader and flatter balance function distributions for all particle species, reflecting a scenario dominated by independent fragmentation and multi-parton interactions, with less pronounced collective dynamics. The observed differences in the near-side peak and overall correlation shapes highlight the important role of the underlying particle production mechanisms, the influence of quark flavor, and the dynamics of hadronization in shaping charge correlations in high-multiplicity proton-proton collisions.

The lack of significant away-side correlation at $\Delta\phi$ $\simeq$ $\pi$ in \pythia is characteristic of color reconnection effects, where local charge/baryon number is conserved but the initial back-to-back momentum correlations are washed out. In contrast, \epos without a core not only reproduces the sharp near-side peak but also preserves a distinct away-side ridge, consistent with a picture of independent string fragmentation where momentum conservation from initial parton scatterings is maintained. Finally, the inclusion of a hydrodynamic core in \epos results in a significant broadening of the balance function, particularly along the $\Delta y$ axis. This smearing is most pronounced for protons, pointing towards delayed hadronization and the effects of collective flow in a thermalized medium, which diffuse the initial correlations over a wider phase space. In the strangeness sector, kaons tend to exhibit narrower balance functions, influenced by local associated $K^{+}K^{-}$ production and the contributions from short-lived resonances such as 
$\phi~\rightarrow~K^{+}K^{-}$. These features result in relatively localized charge correlations, with kaons showing less sensitivity to longitudinal broadening than baryons, and they offer complementary insight into strangeness production and hadronization dynamics.
\subsection{One-dimensional projections}

\begin{figure*}[!bth]  
    \centering
    \subfigure[]{
        \includegraphics[width=0.3\textwidth]{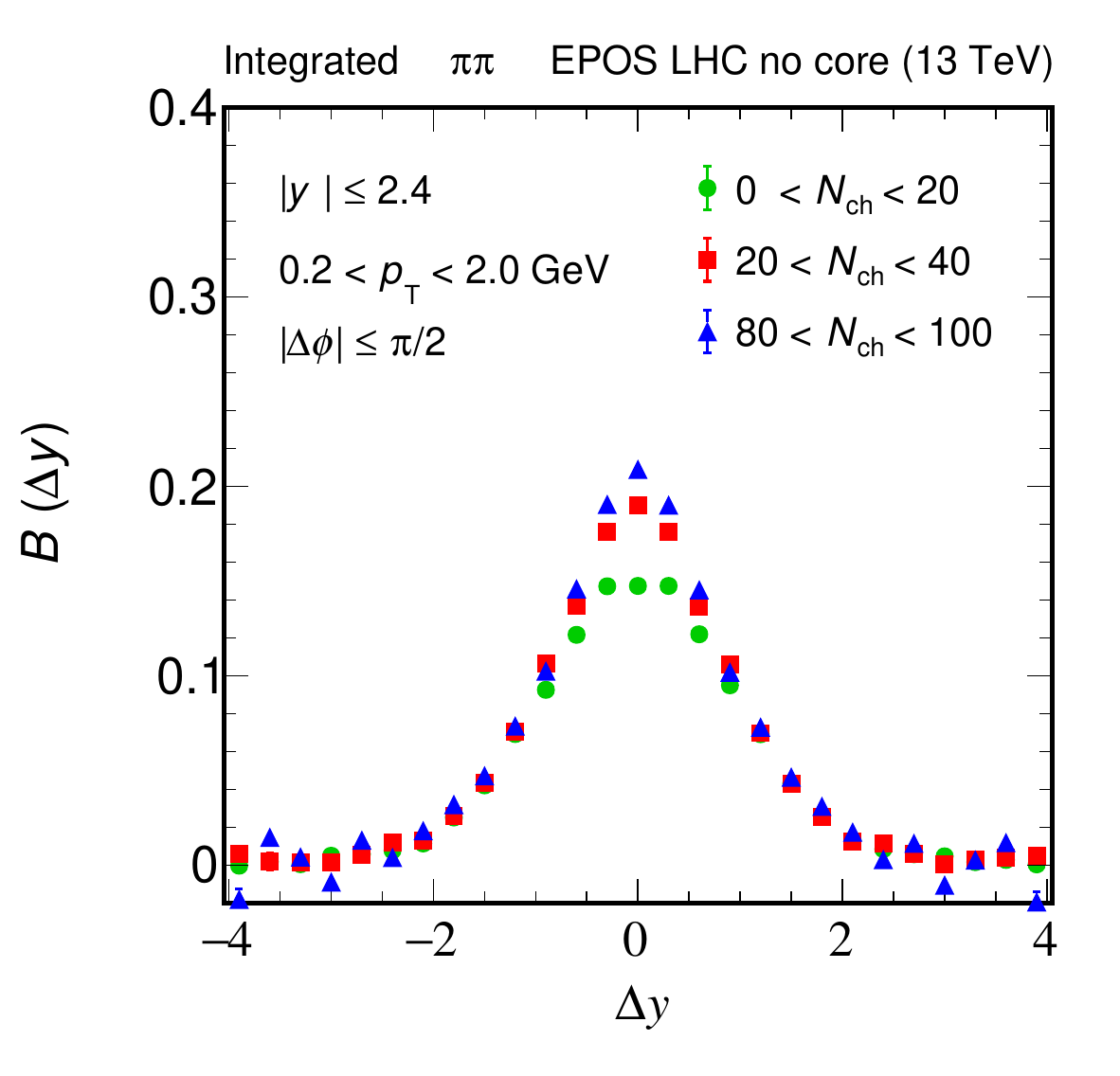}
    }
    \subfigure[]{
        \includegraphics[width=0.3\textwidth]{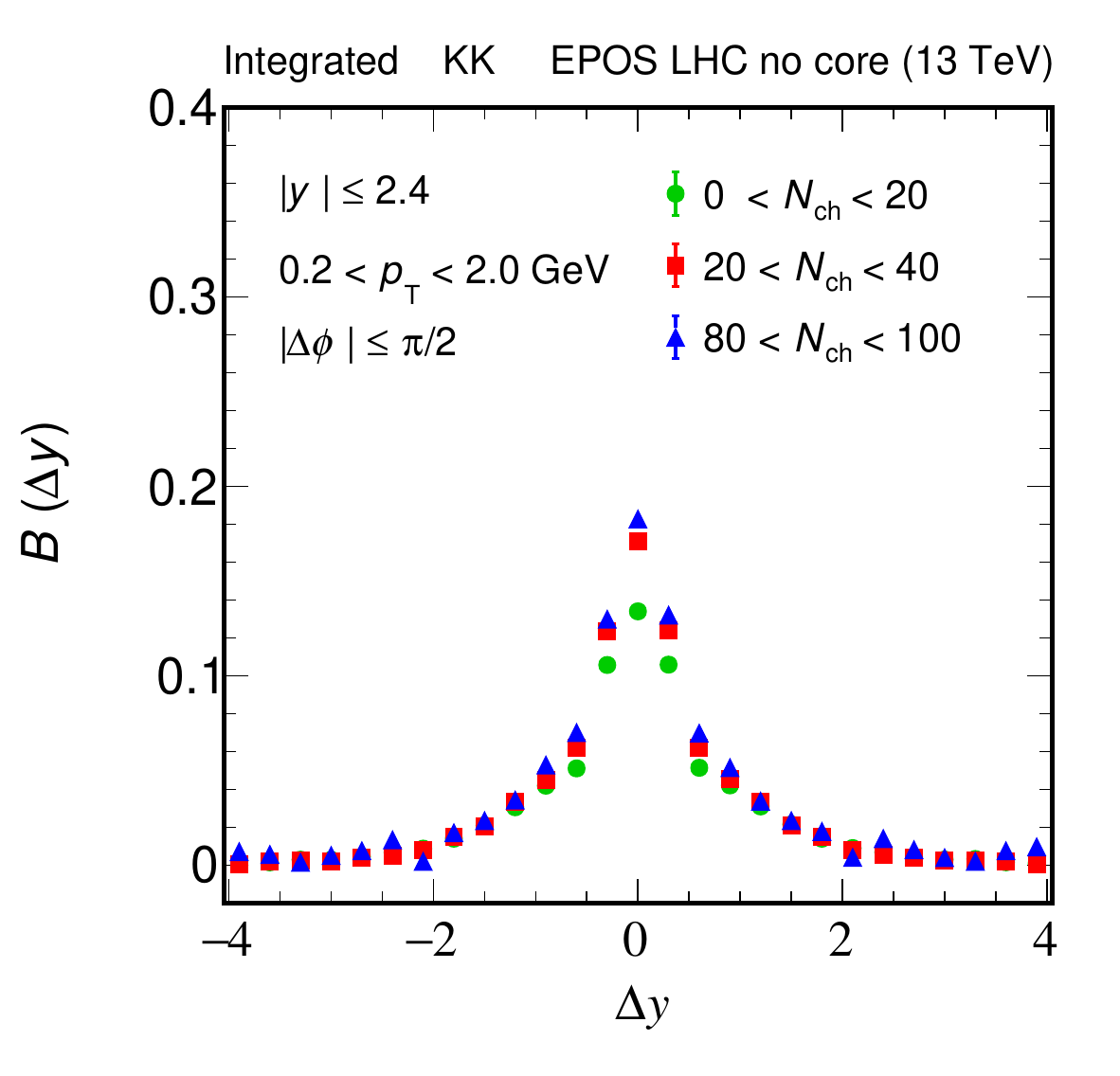}
    }
    \subfigure[]{
        \includegraphics[width=0.3\textwidth]{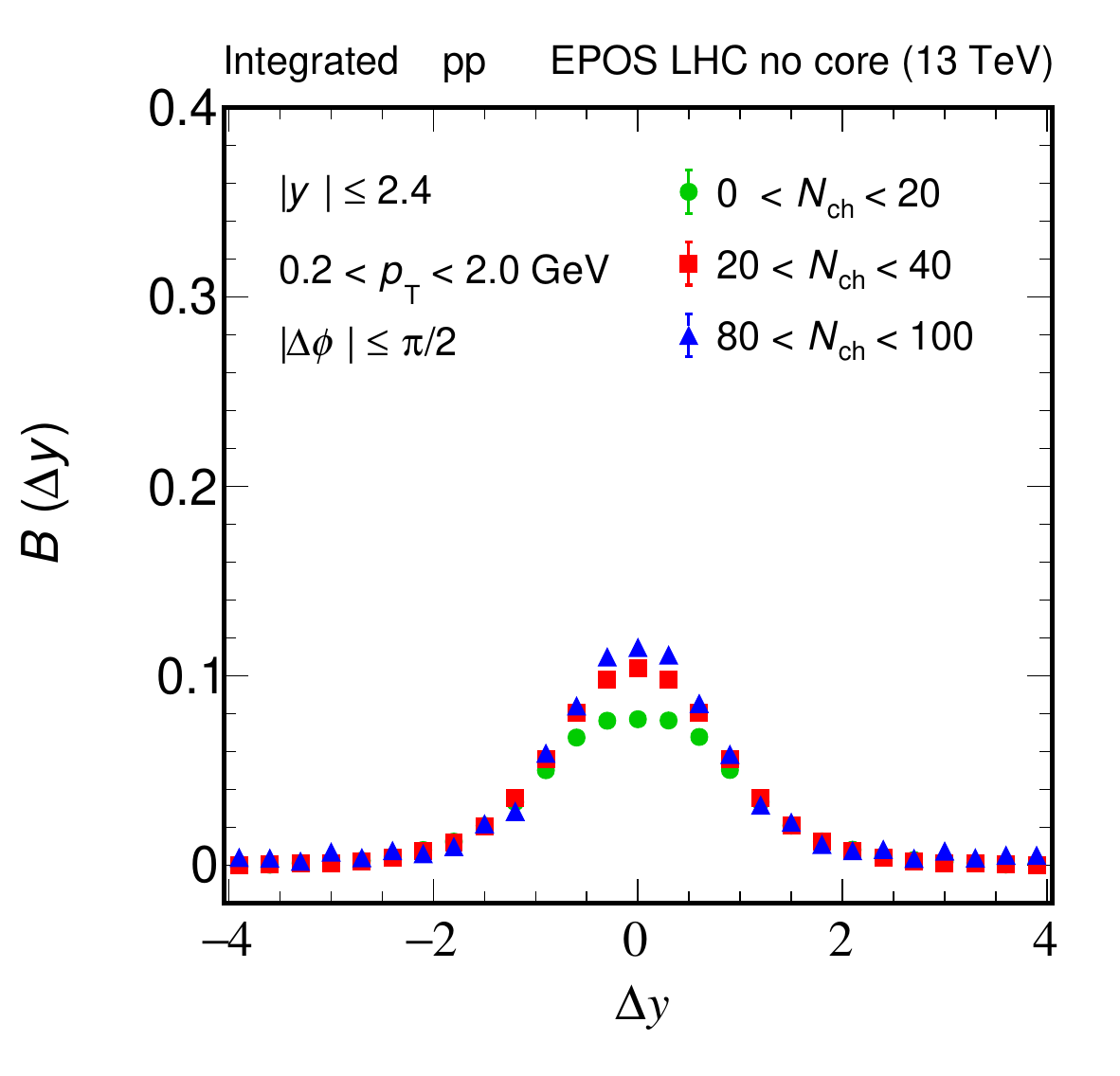}
    }
    \subfigure[]{
        \includegraphics[width=0.3\textwidth]{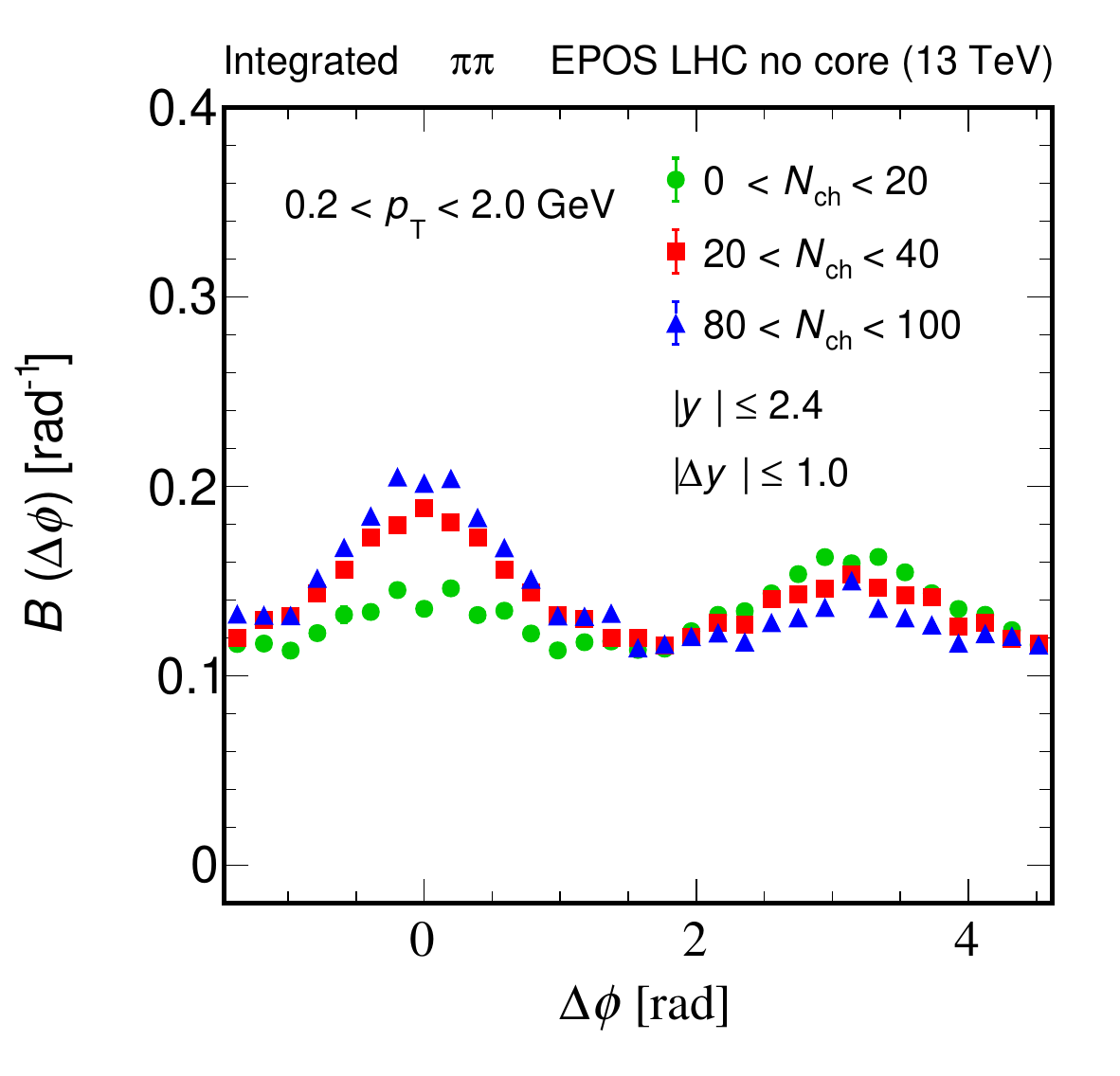}
    }
    \subfigure[]{
        \includegraphics[width=0.3\textwidth]{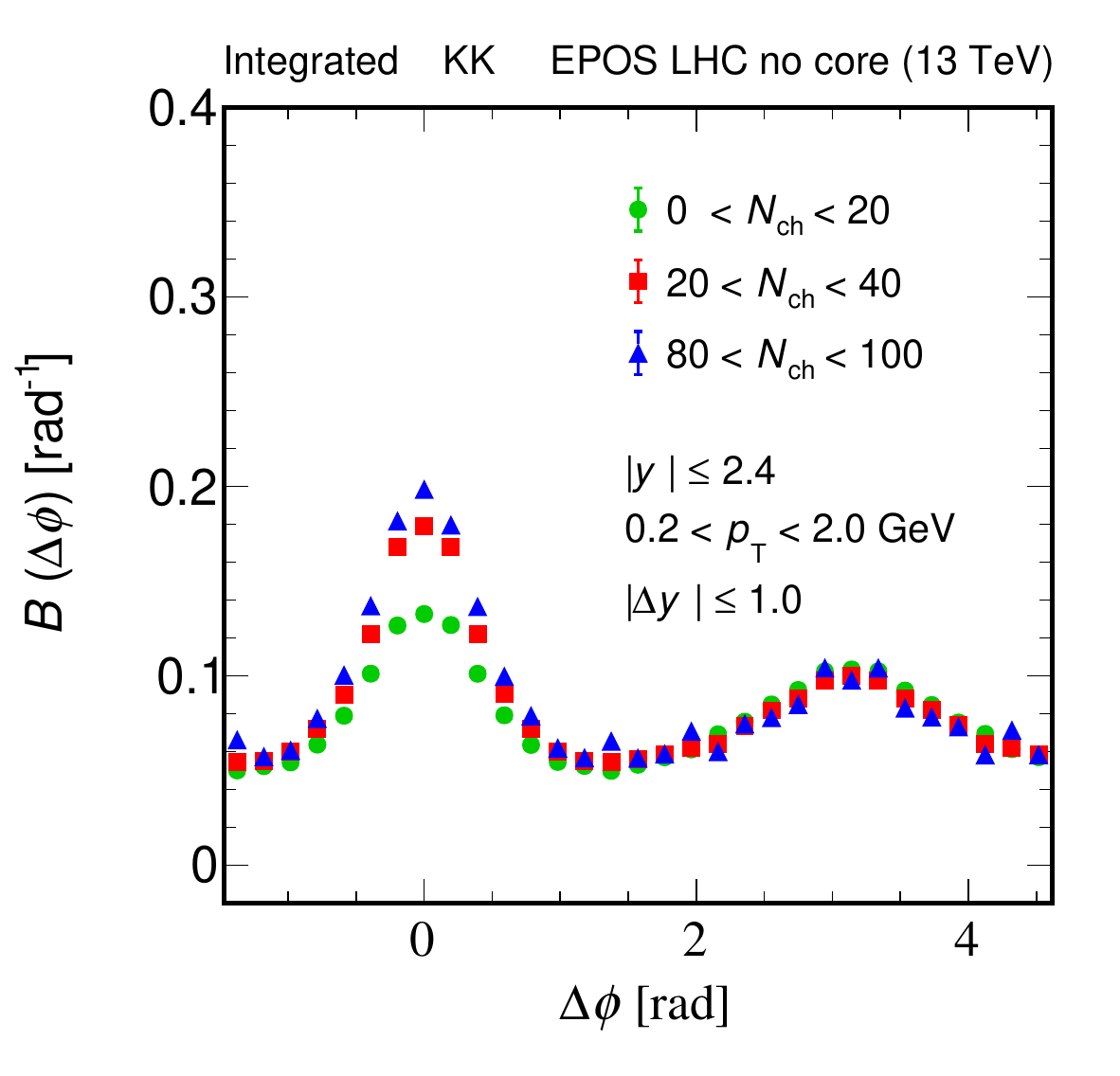}
    }
    \subfigure[]{
        \includegraphics[width=0.3\textwidth]{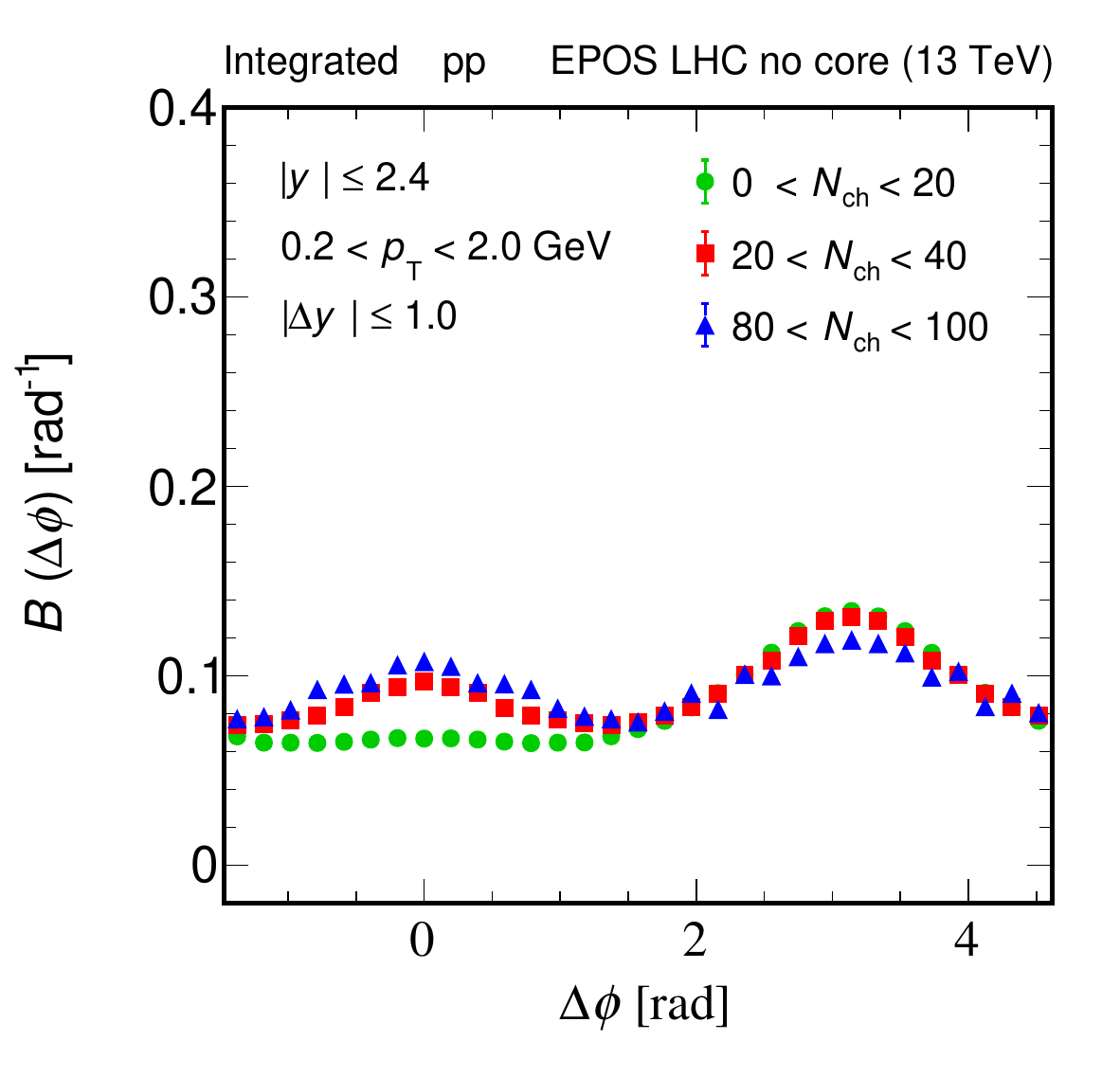}
    }
 \caption{One-dimensional projections of $B$ along $\Delta y$ and $\Delta\phi$ from \epos model without core in pp collisions at $\sqrt{s} =$ 13 TeV. The left column is for pion, the middle column is for kaon, and the right panel is for proton, respectively. $\Delta\phi$ projections are take in $|\Delta\phi| \leq \pi/2$ and $\Delta\phi$ projections are taken in $|\Delta y| \leq 1.0$ range.}
    \label{fig:model_plots_1d_epos_nc}
\end{figure*}

\begin{figure*}[!bth]  
    \centering
    \subfigure[]{
        \includegraphics[width=0.3\textwidth]{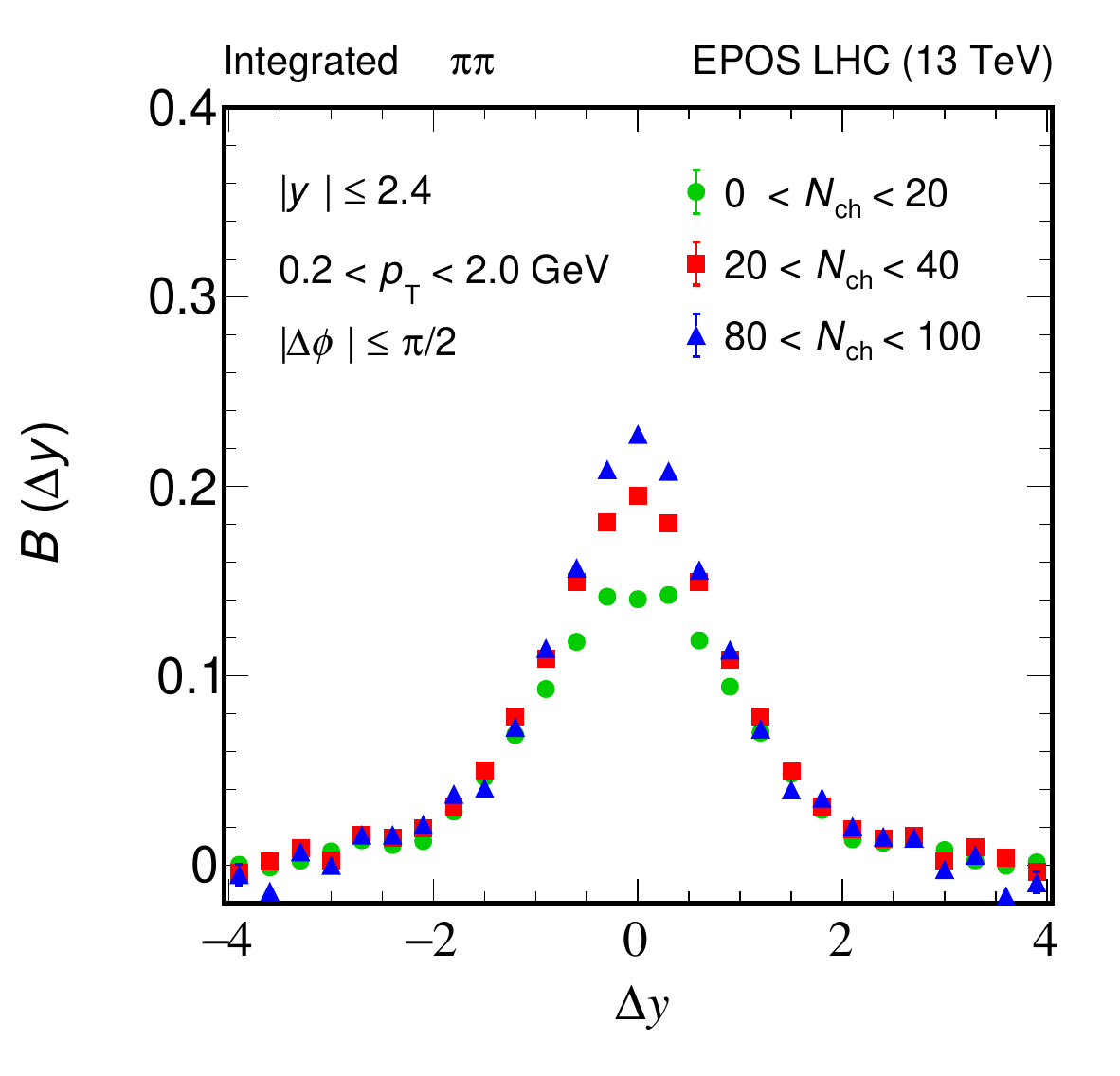}
    }
    \subfigure[]{
        \includegraphics[width=0.3\textwidth]{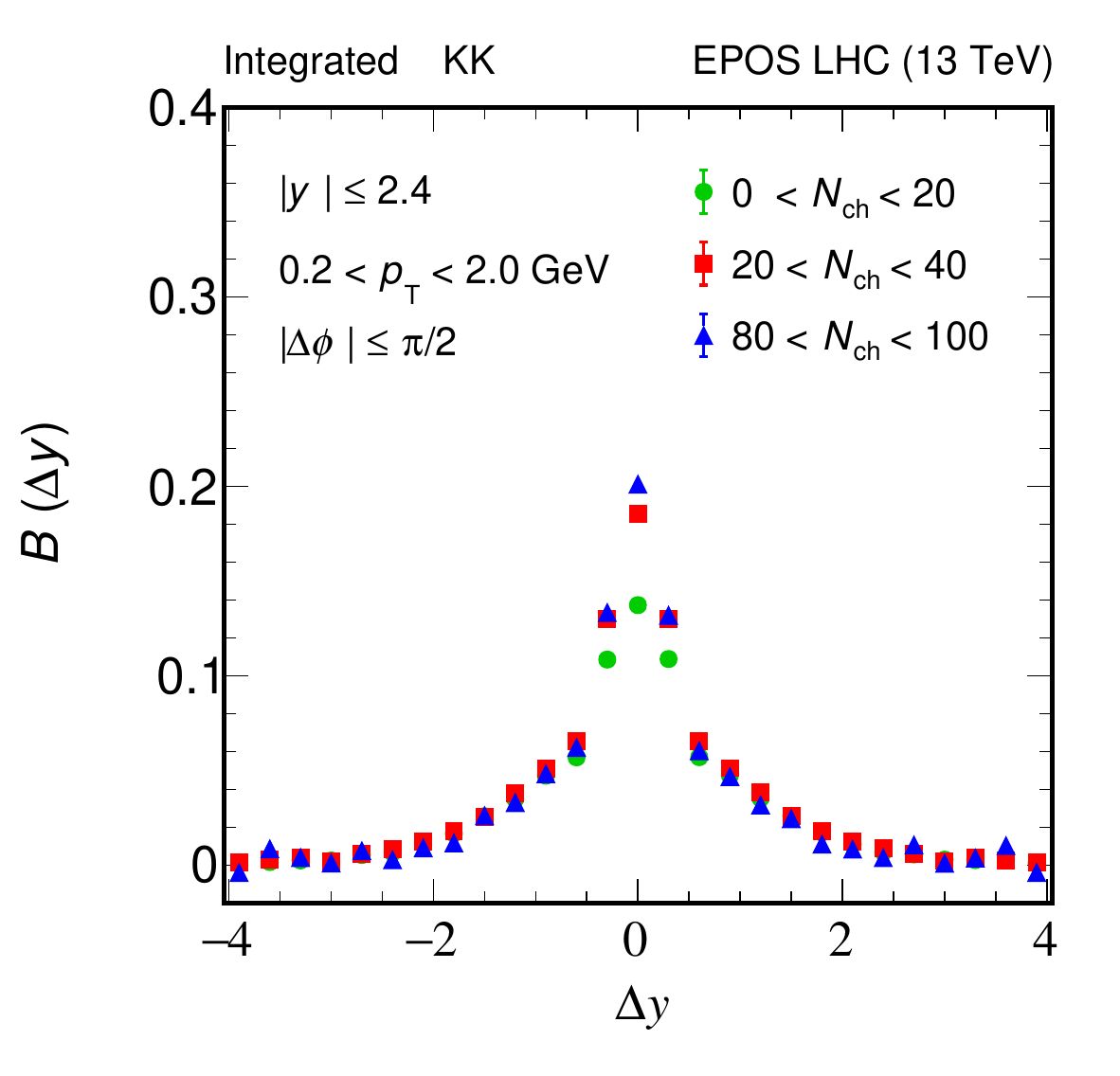}
    }
    \subfigure[]{
        \includegraphics[width=0.3\textwidth]{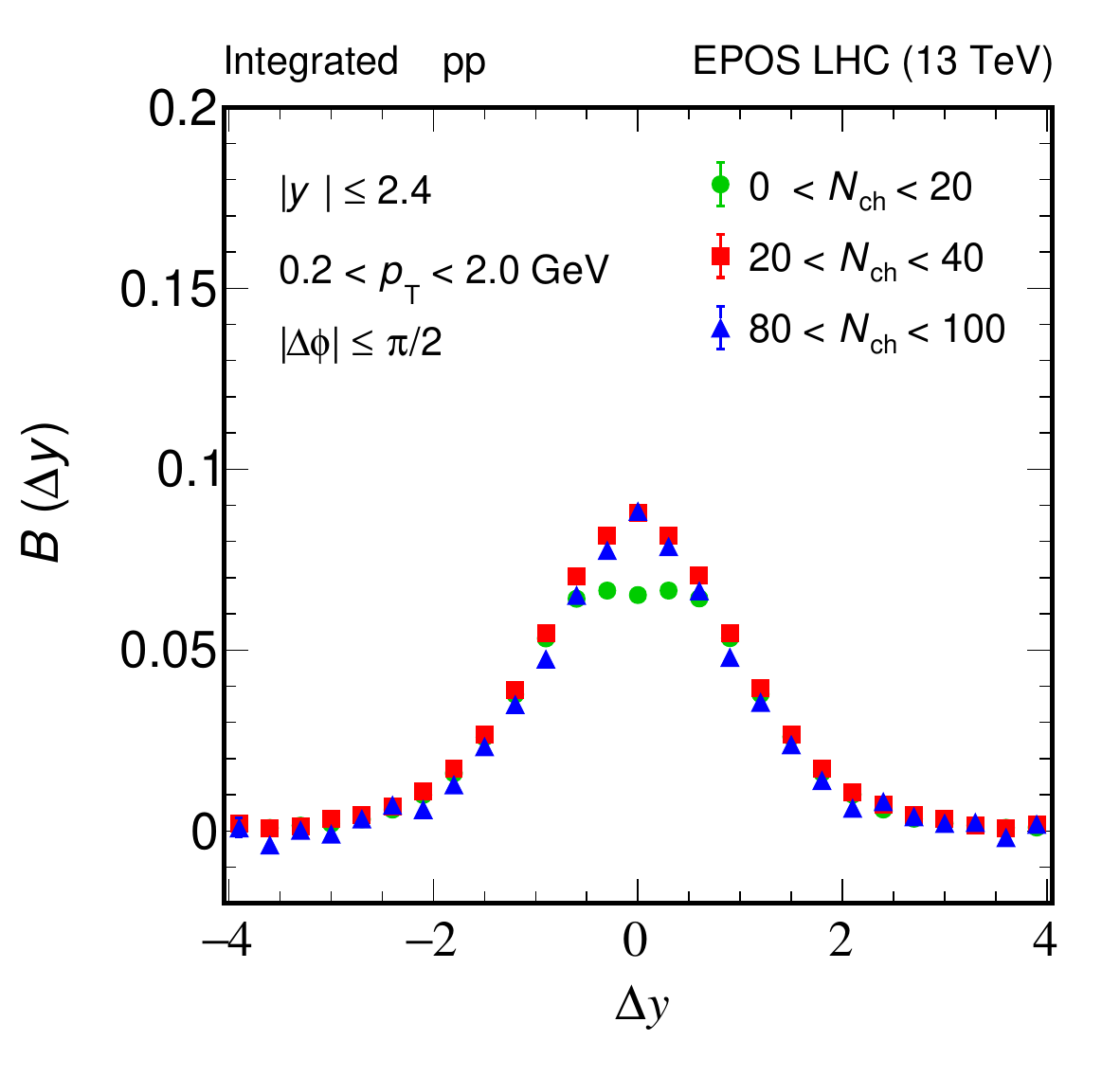}
    }
    \subfigure[]{
        \includegraphics[width=0.3\textwidth]{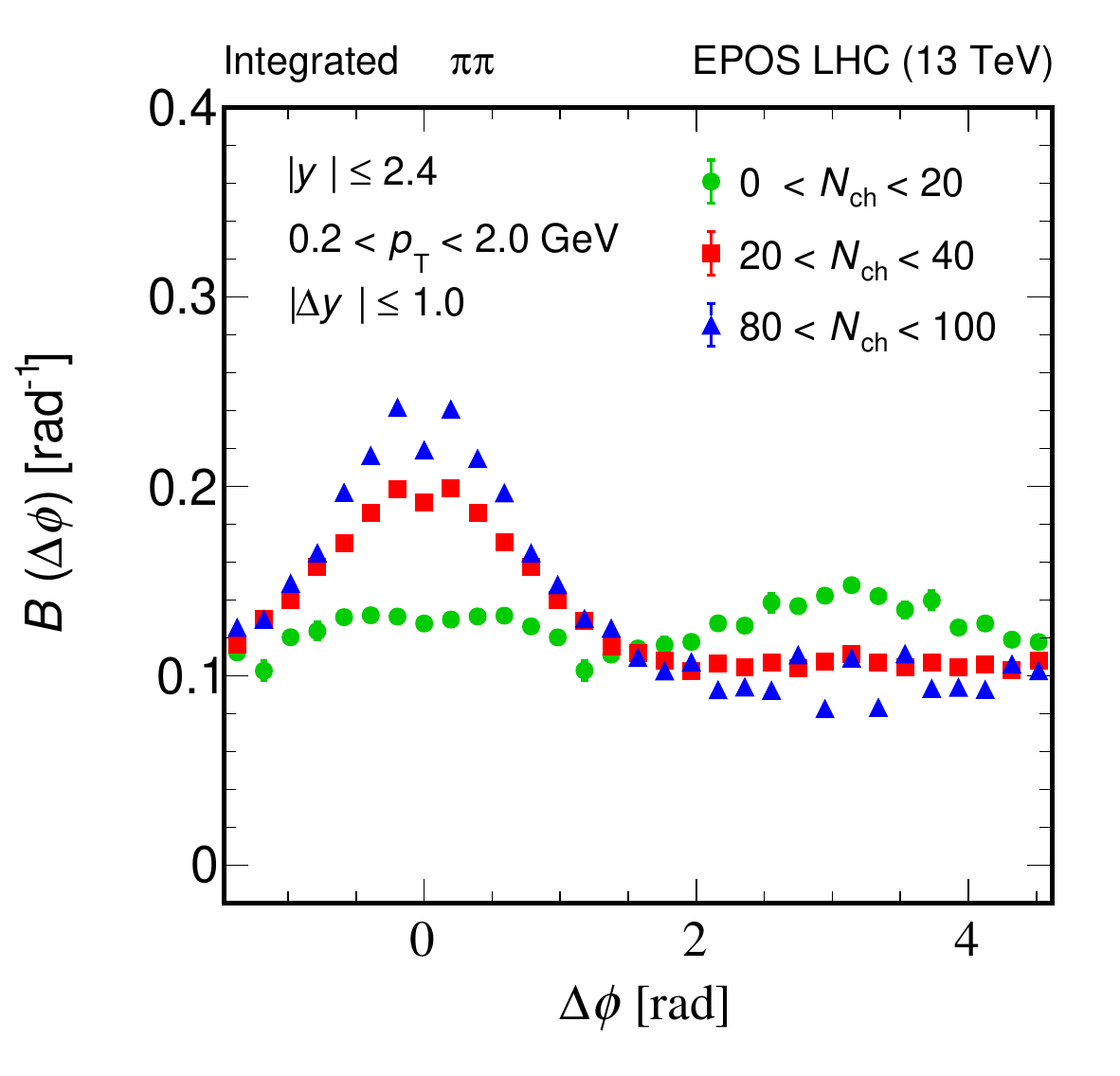}
    }
    \subfigure[]{
        \includegraphics[width=0.3\textwidth]{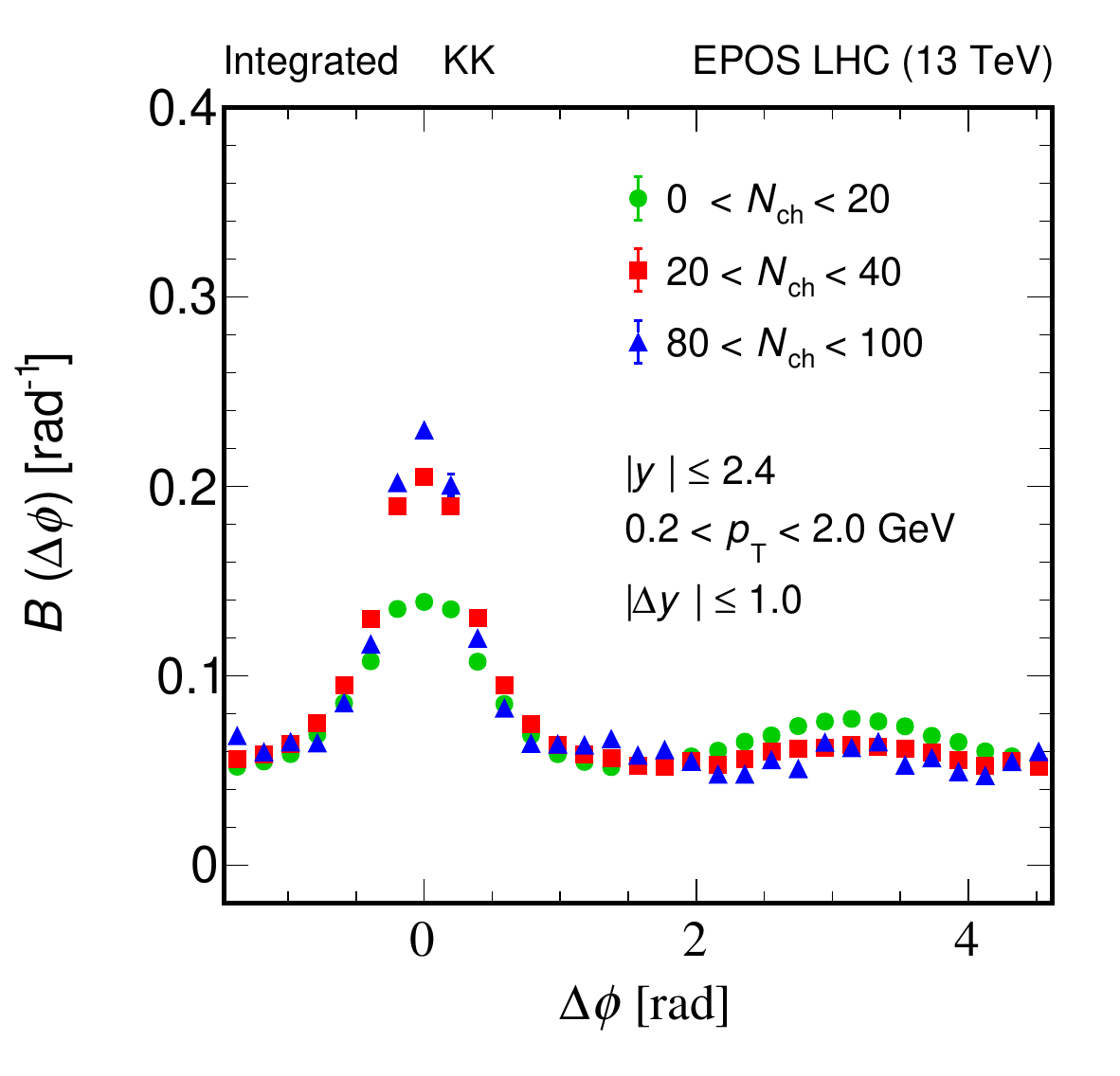}
    }
    \subfigure[]{
        \includegraphics[width=0.3\textwidth]{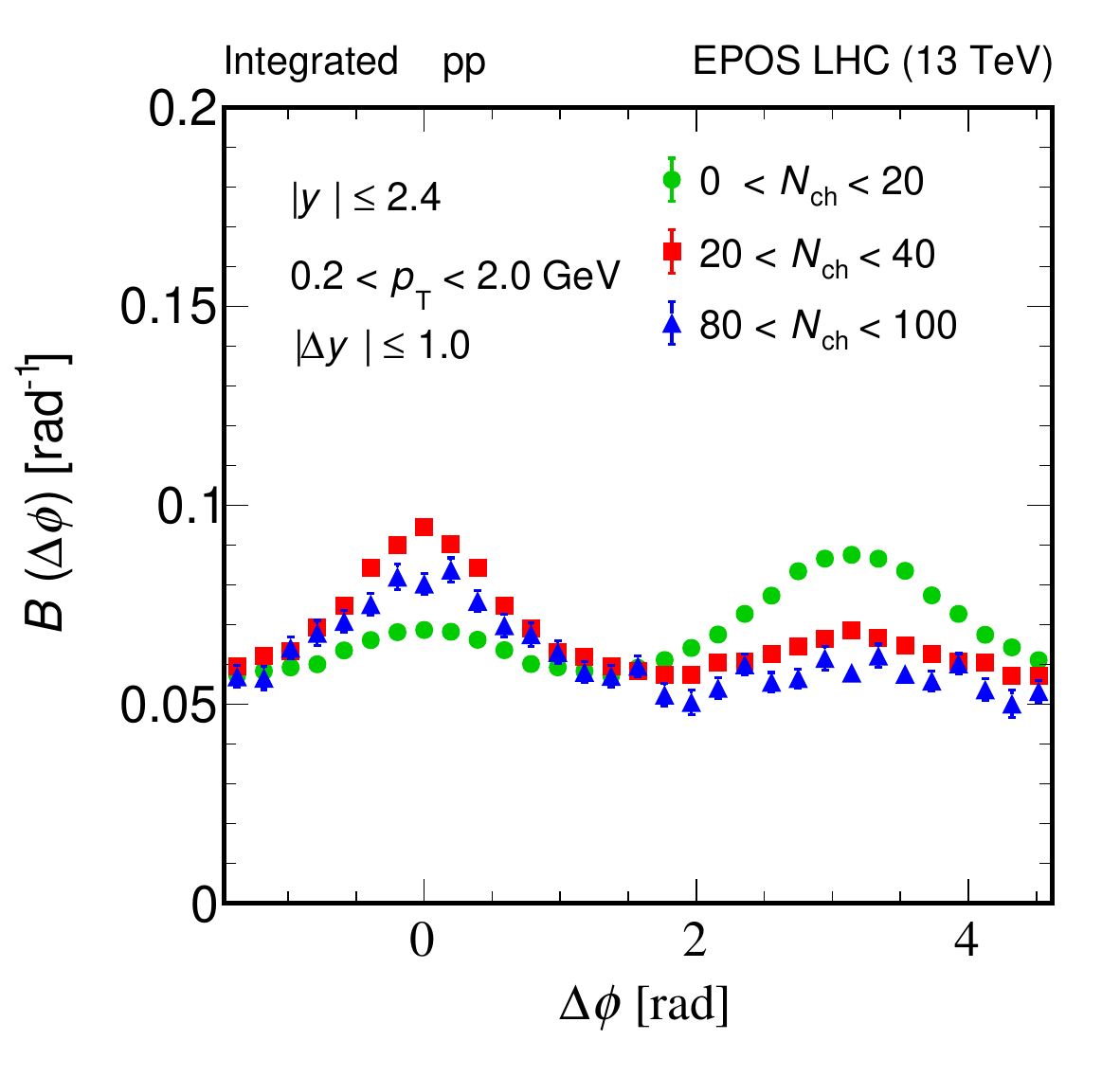}
    }
 \caption{One-dimensional projections of $B$ along $\Delta y$ and $\Delta\phi$ from \epos model  in pp collisions at $\sqrt{s} =$ 13 TeV. The left column is for pion, the middle column is for kaon, and the right panel is for proton, respectively. $\Delta y$ projections are take in $|\Delta\phi| \leq \pi/2$ and $\Delta\phi$ projections are taken in $|\Delta y| \leq 1.0$ range.}
    \label{fig:model_plots_1d_epos}
\end{figure*}
Figure~\ref{fig:model_plots_1d}, \ref{fig:model_plots_1d_epos_nc} and  \ref{fig:model_plots_1d_epos} present the one-dimensional projections of the balance functions from \pythia and \epos simulations without and  with core inclusion for different identified hadron species over three multiplicity classes. The high-multiplicity class corresponds to ( $ 80 < N_{\mathrm{ch}} < 100$), while \( \langle N_{\mathrm{ch}}\rangle = 8 \) represents low-multiplicity events. The top row (panels a--c) presents the balance function \( B(\Delta y) \), while the bottom row (panels d--f) shows \( B(\Delta\phi) \). The projections are performed along \( \Delta y \) (within \( \Delta\phi \leq \pi/2 \)) and \( \Delta\phi \) (within \( |\Delta y| \leq 1.0 \)),  considering particles within the kinematic range \( |y| \leq 2.4 \) and \( 0.2 < p_{\mathrm{T}} < 2.0\ \mathrm{GeV} \), integrated over all spherocity intervals. 

In \pythia simulations, the $\Delta\phi$ balance function for pions, kaons, and protons evolves noticeably with multiplicity. At low multiplicity, in addition to near-side peak, a strong away-side peak appears near $\Delta\phi \sim \pi$, reflecting back-to-back jet fragmentation. As multiplicity increases, a clear near-side peak develops for all species, while the away-side peak broadens and weakens. This trend highlights the increasing role of soft processes and local charge conservation in high-multiplicity events, which enhances near-side correlations and reduces the dominance of jet-like production~\cite{Lonnblad:2023kft}. In contrast, \epos features both independent string fragmentation in the corona and collective hydrodynamic expansion in the core. When the hydrodynamic core is included, high-multiplicity events show a narrowing of the balance function in $\Delta\phi$, due to the strong collimating effect of radial flow, which drives balancing partners closer together in azimuth. At the same time, the $\Delta y$ balance function broadens as longitudinal flow and diffusion separate charges more widely in rapidity. If the hydrodynamic core is switched off (the “corona-only” scenario), \epos behaves similarly to \pythia, producing broader $\Delta\phi$ distributions. This contrast highlights that the presence of a hydrodynamic core leads to distinct narrowing in azimuth and broadening in rapidity, signatures of collective expansion absent in models based purely on independent fragmentation.
Given the large number of combinations of particle species, multiplicity intervals, and spherocity classes, we do not show the full set of two-dimensional correlation plots or all possible one-dimensional projections. Instead, to provide a concise and informative summary, we focus on the root-mean-square (RMS) widths of the balance functions. In the subsequent discussion, these widths are compared for the spherocity-integrated events as well as for the top 20\% (isotropic) and bottom 20\% (jet-like) event classes. This methodology offers a clear way to highlight the influence of event topology on charge-dependent correlations in both \pythia and \epos.


\begin{figure*}[!htb]  
    \centering
    \subfigure[]{
        \includegraphics[width=0.3\textwidth]{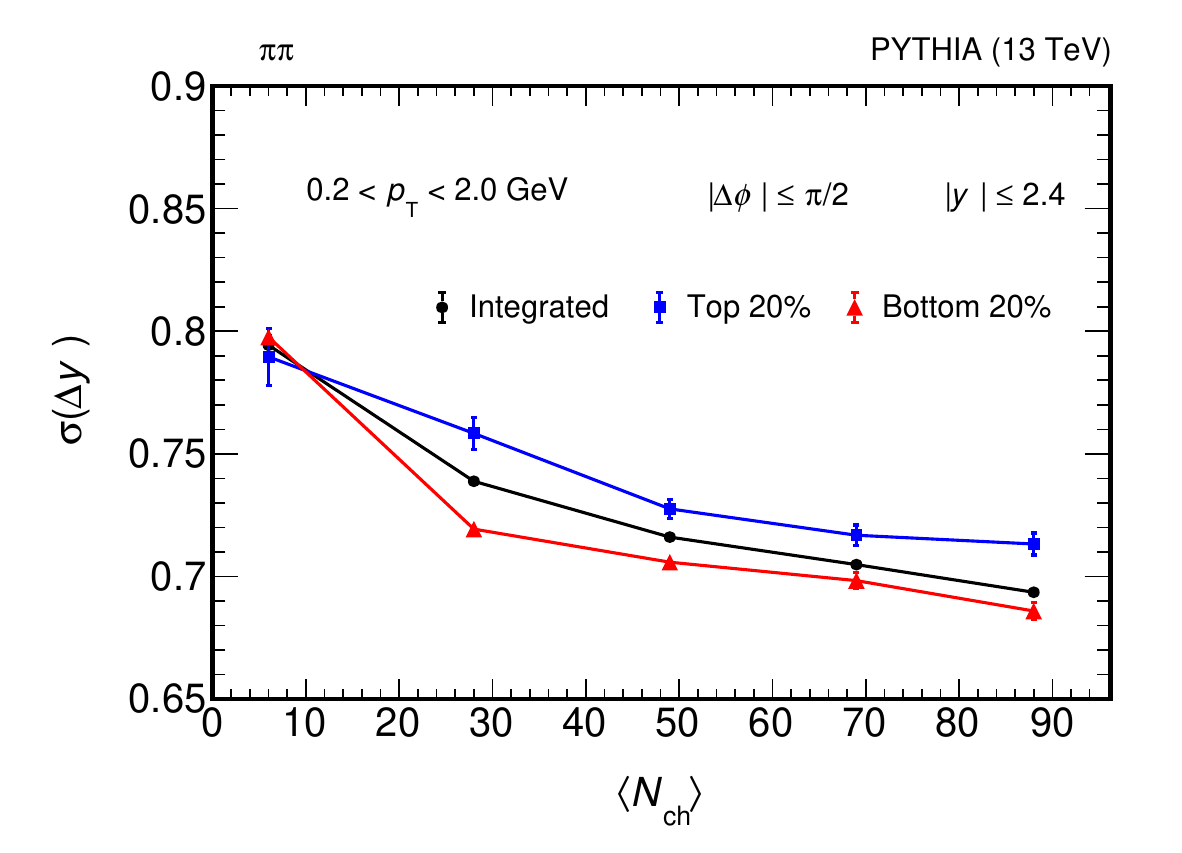}
    }
    \subfigure[]{
        \includegraphics[width=0.3\textwidth]{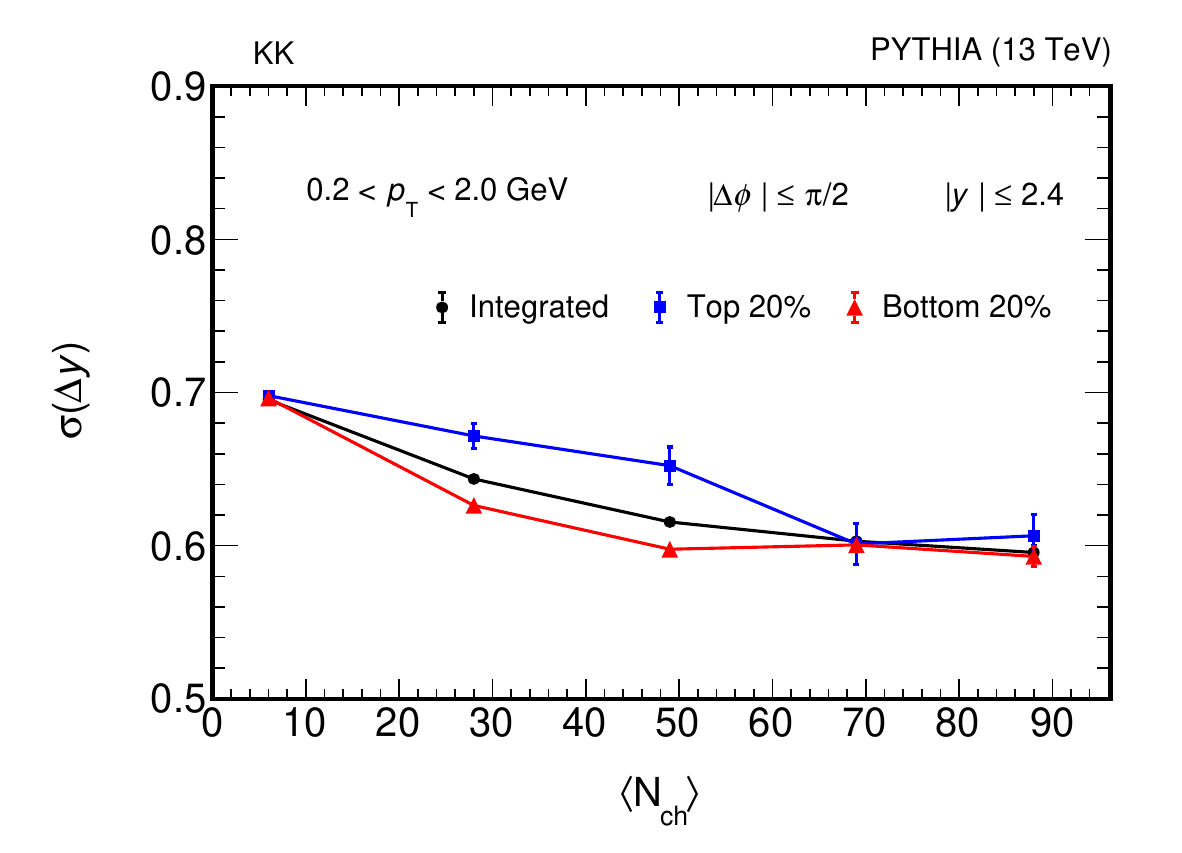}
    }
    \subfigure[]{
        \includegraphics[width=0.3\textwidth]{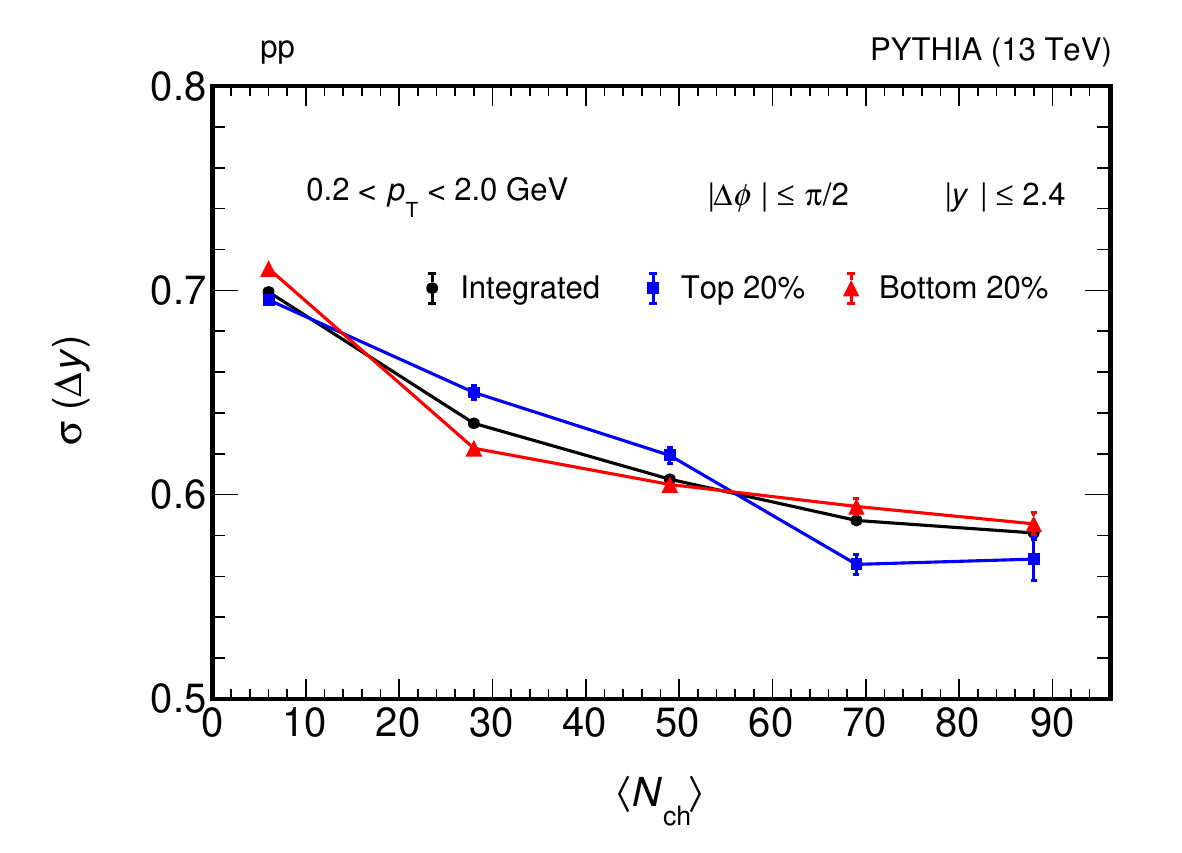} 
    }

    \subfigure[]{
        \includegraphics[width=0.3\textwidth]{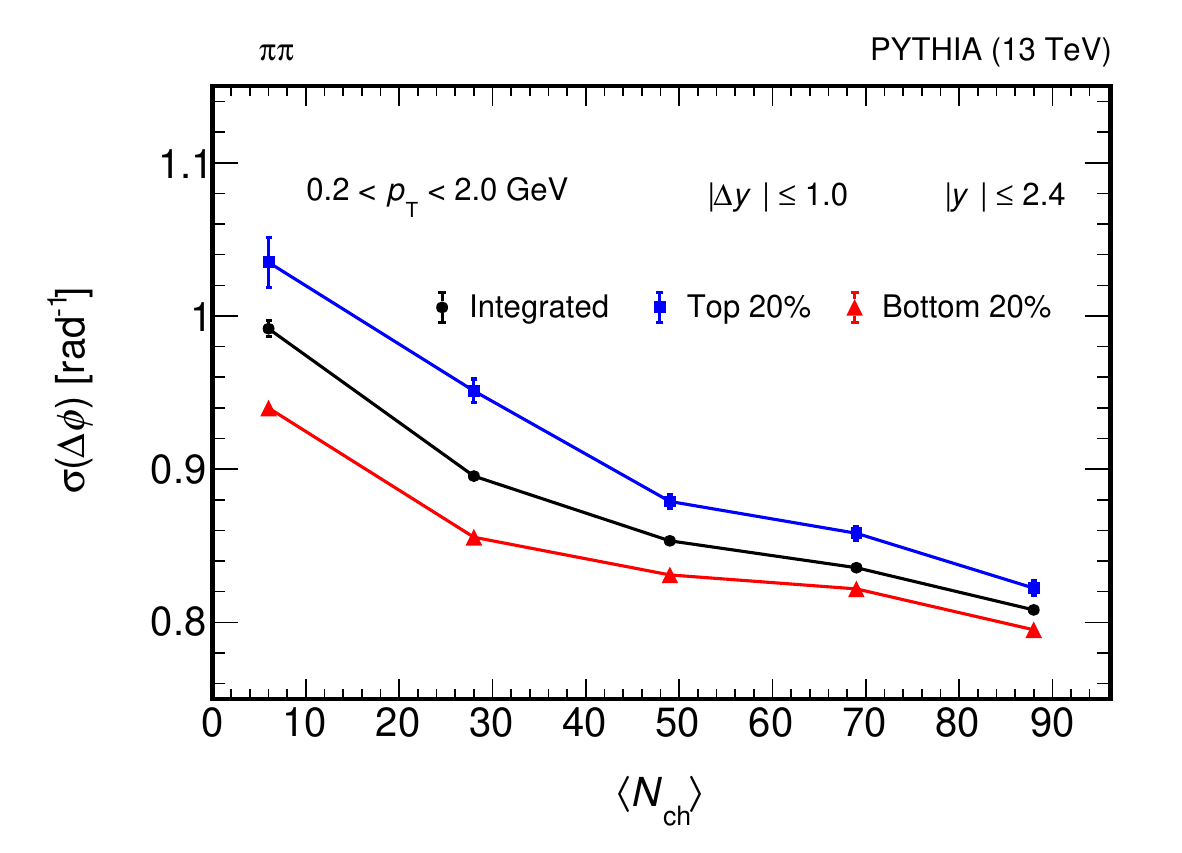}
    }
    \subfigure[]{
        \includegraphics[width=0.3\textwidth]{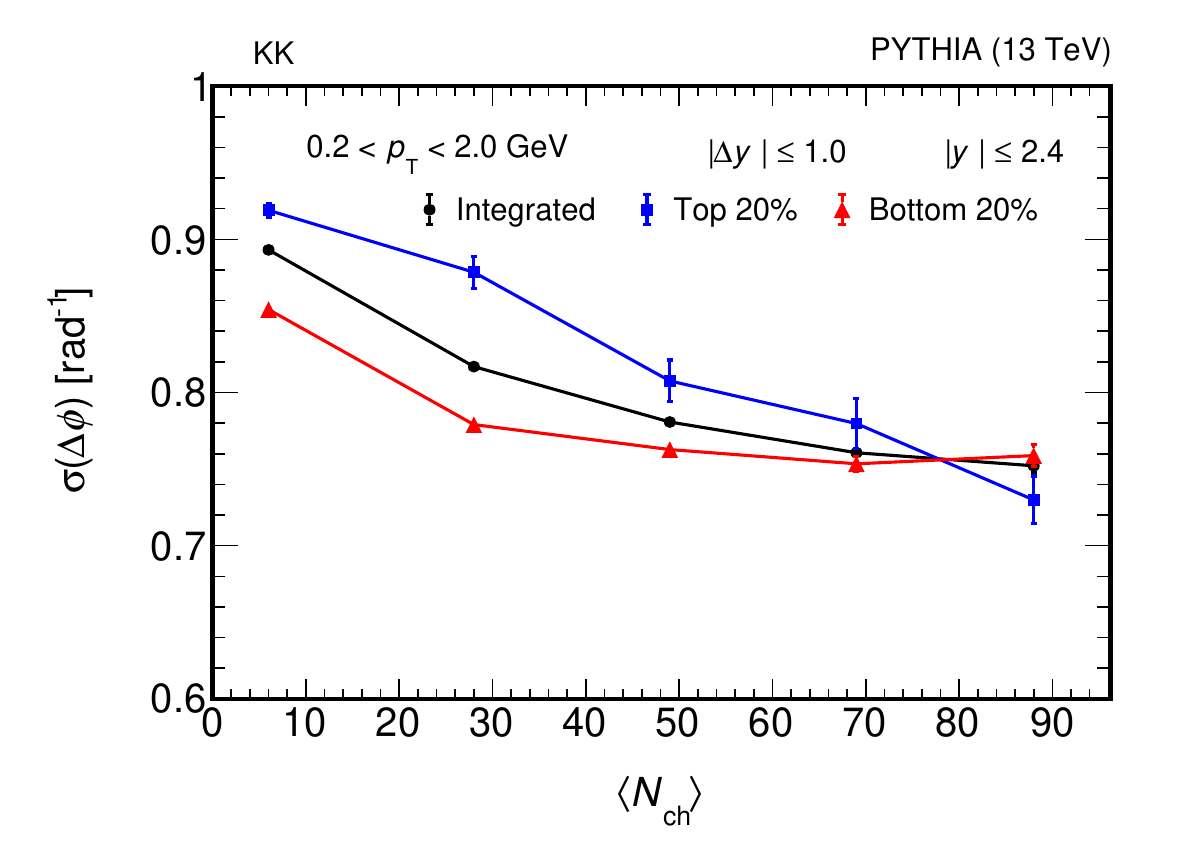}
    }
    \subfigure[]{
        \includegraphics[width=0.3\textwidth]{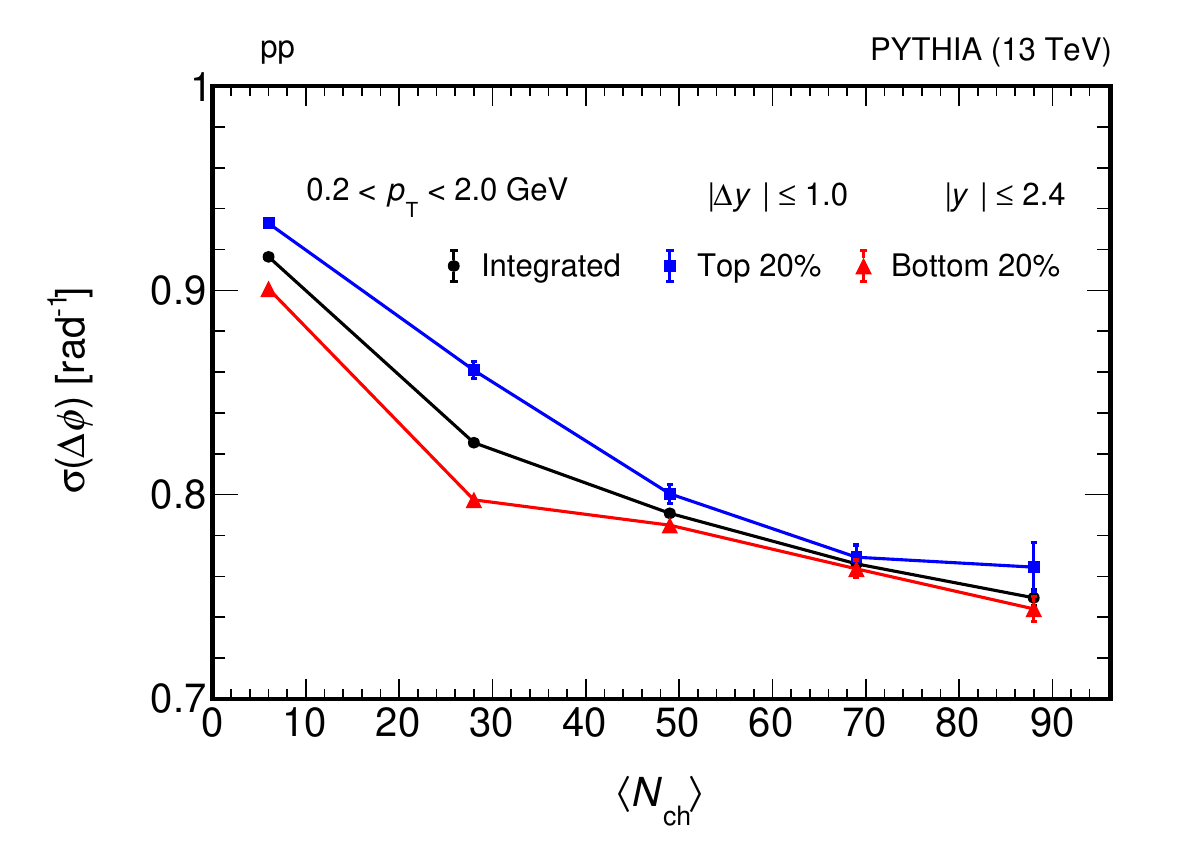} 
    }
    
\caption{The width of $B$ for $\pi$ (left panel), $K$ (right panel) and $p$ (middle panel) in pp collisions at $\sqrt{s} = $13 TeV as a function of $N_\mathrm{ch}$ and spherocity in \pythia model. The plot in the upper panel is for $\Delta y$ width and the lower panel is for $\Delta\phi$ width calculation.}
    \label{fg:widthModel1}
\end{figure*}
\subsection{Balance functions width}

\begin{figure*} [!htb] 
    \centering
    \subfigure[]{
        \includegraphics[width=0.3\textwidth]{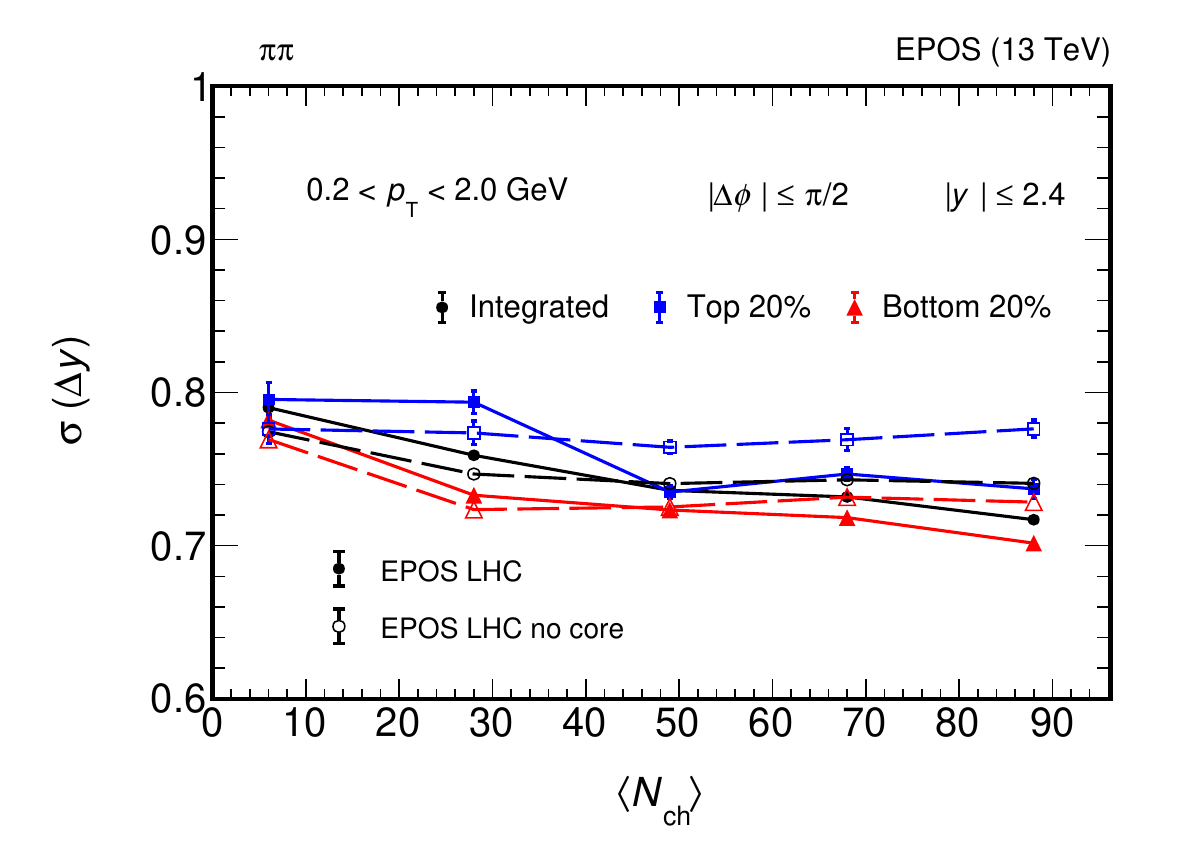} 
        
    }
    \subfigure[]{
        \includegraphics[width=0.3\textwidth]{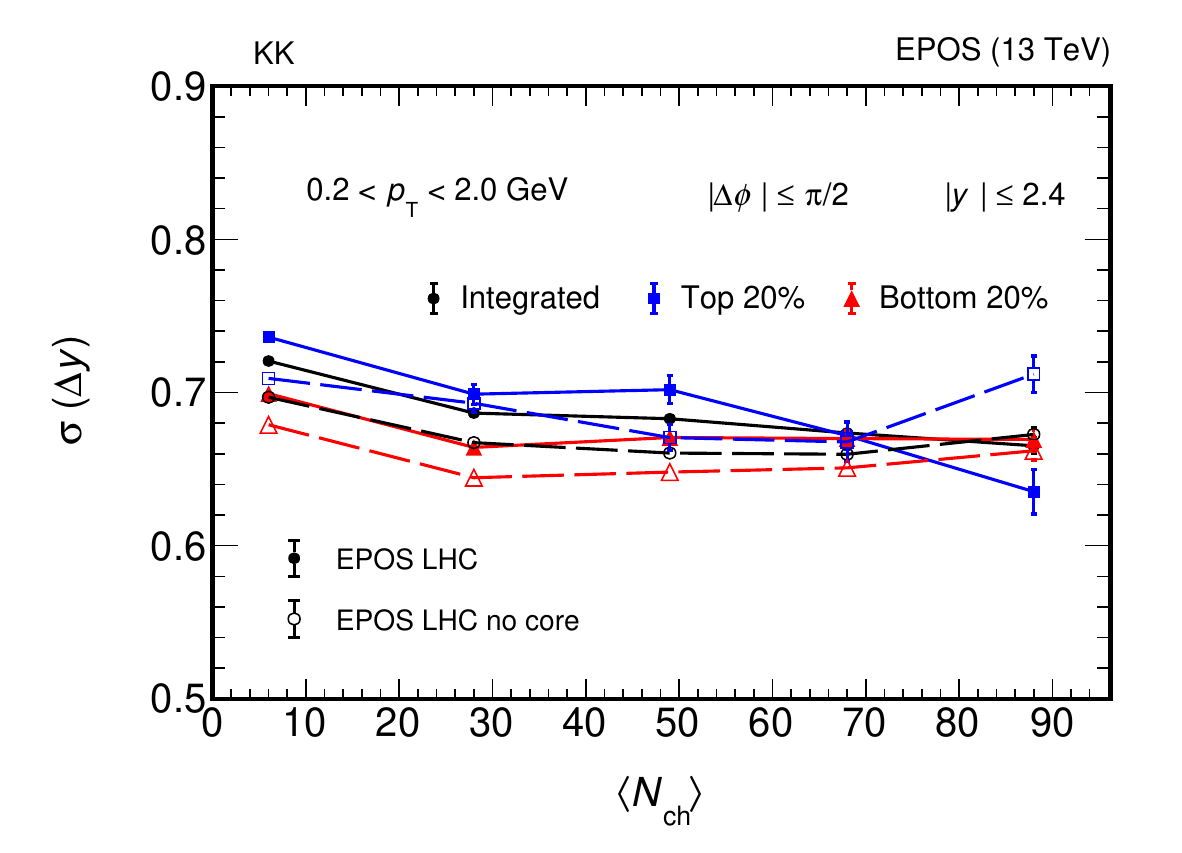} 
    }
    \subfigure[]{
        \includegraphics[width=0.3\textwidth]{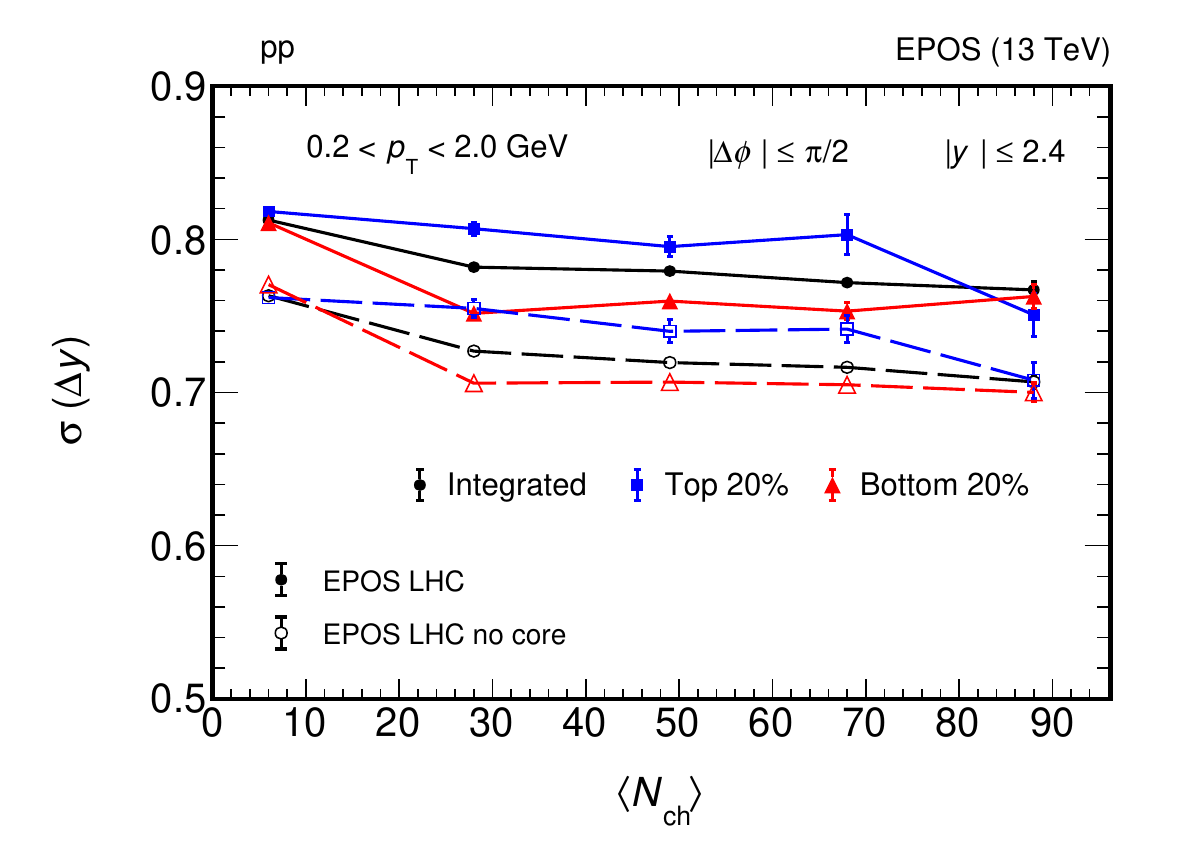} 
        
    }
    \subfigure[]{
        \includegraphics[width=0.3\textwidth]{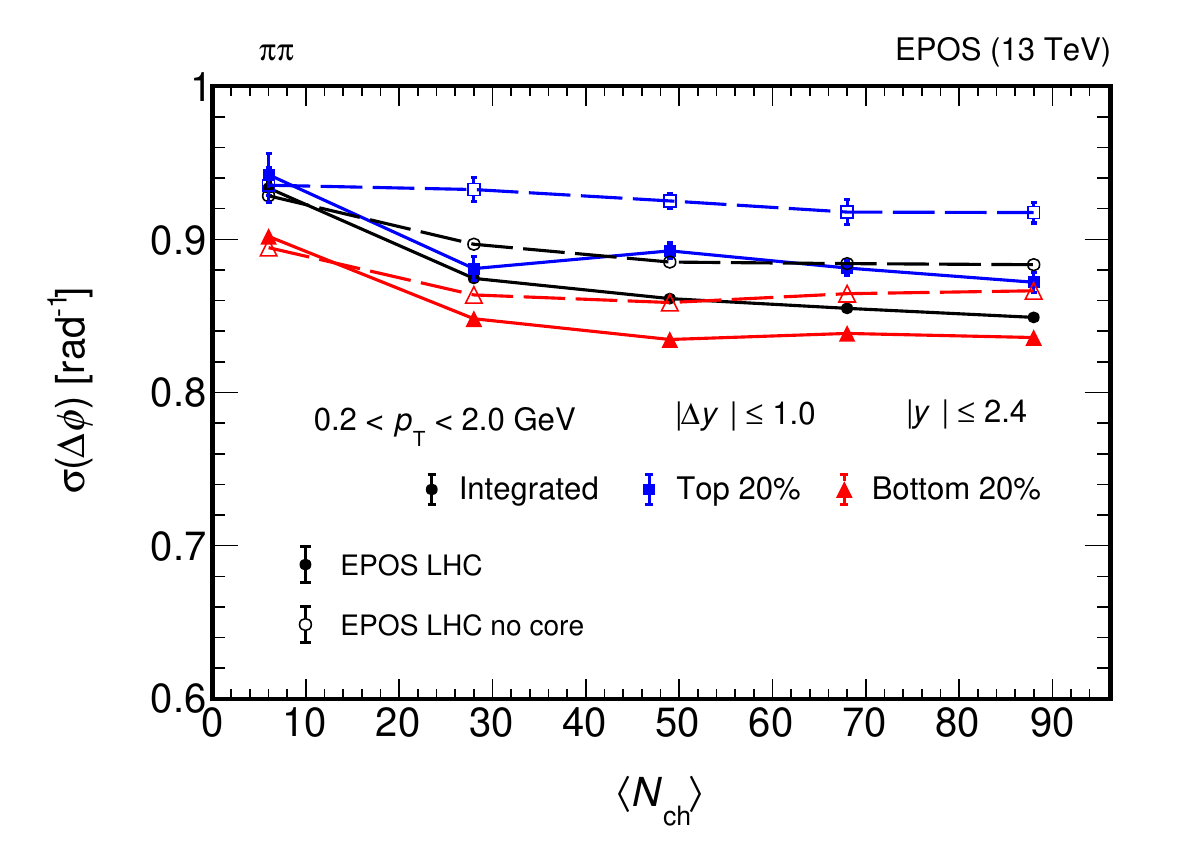} 
        
    }    
    \subfigure[]{
        \includegraphics[width=0.3\textwidth]{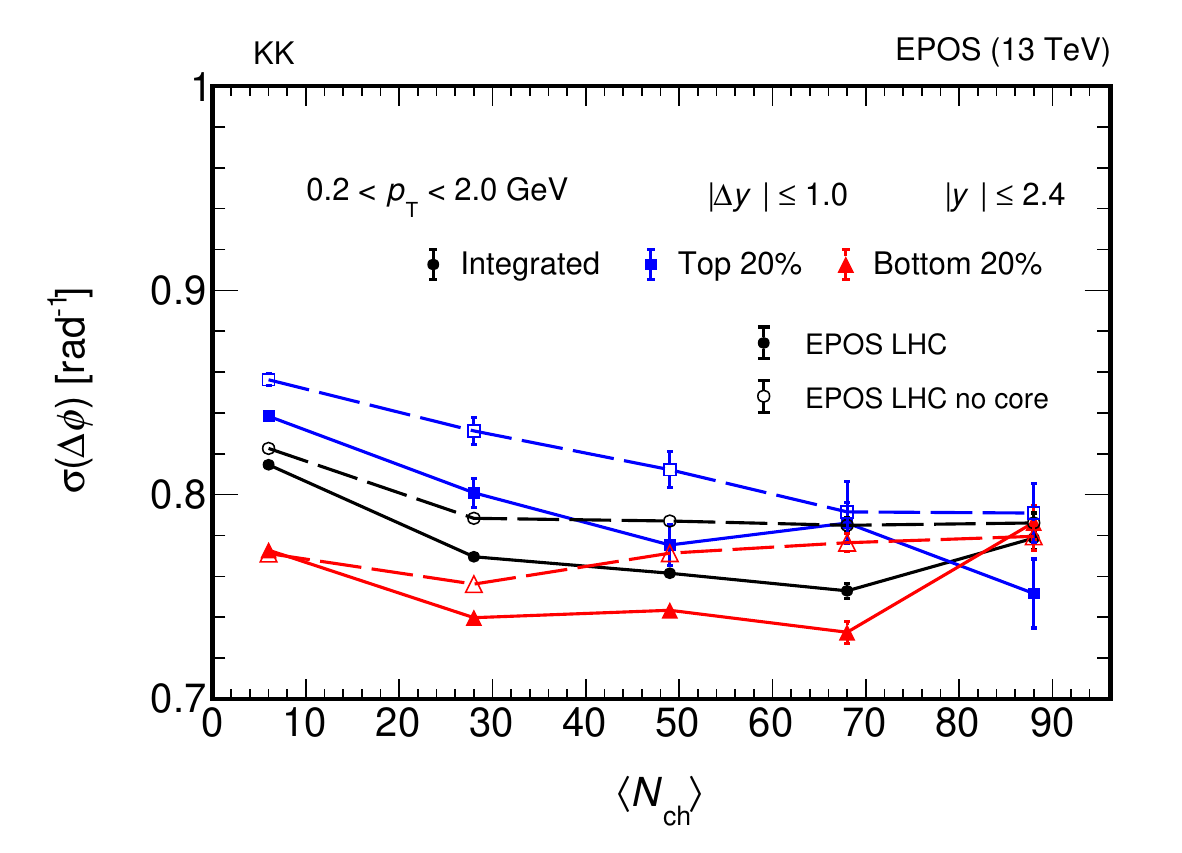} 
        }
    \subfigure[]{
        \includegraphics[width=0.3\textwidth]{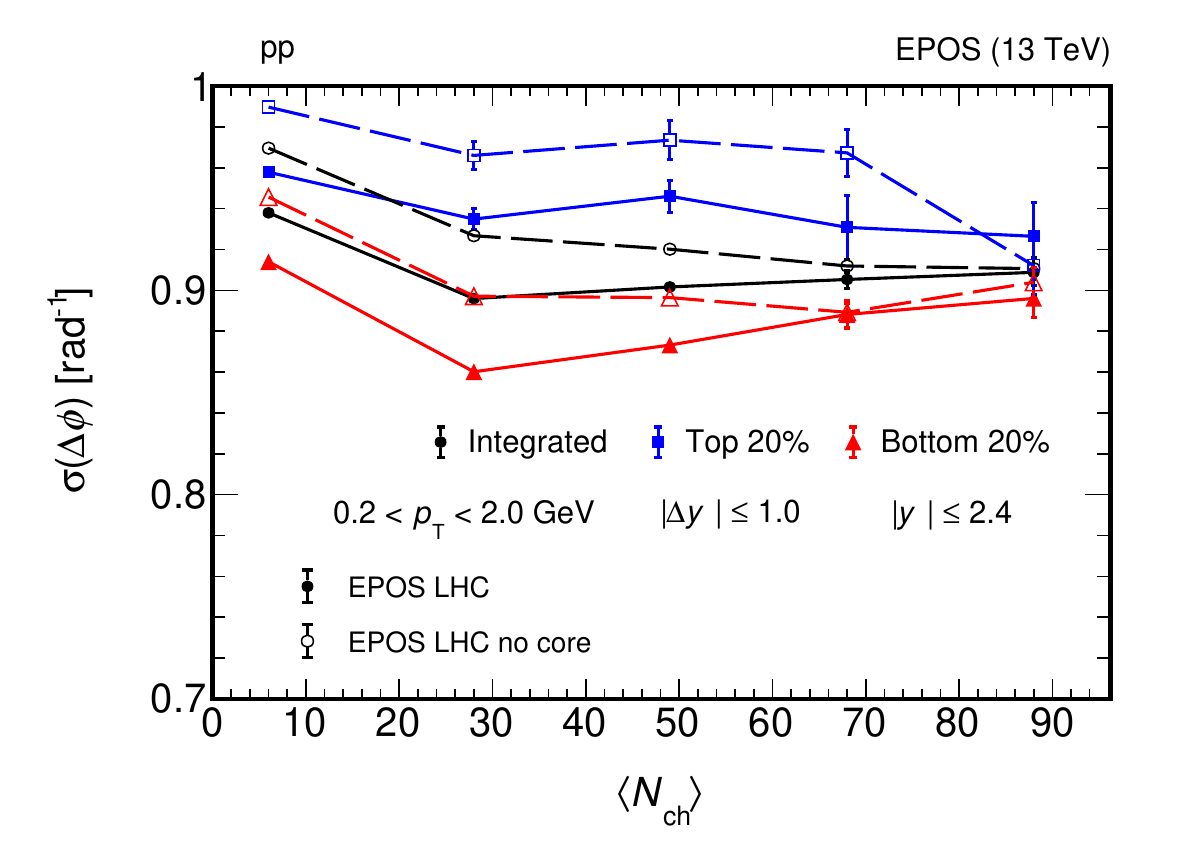} 
        
    }
    
\caption{The width of $B$ for $\pi, K$ and $p$ in pp collisions at $\sqrt{s} = $13 TeV as a function of $N_\mathrm{ch}$ and spherocity in \epos model. The plot in the upper panel is for $\Delta y$ width and the lower panel is for $\Delta\phi$ width calculation.}
\label{fg:eposwidthModel}
\end{figure*}

Before presenting the balance-function widths, we note that the multiplicity axis in the width figures corresponds to the average charged-particle multiplicity $\langle N_{\mathrm{ch}} \rangle$ of each selected multiplicity class. The evolution of balance function widths is summarized for \pythia in Figure~\ref{fg:widthModel1} and for \epos in Figure~\ref{fg:eposwidthModel}, where the latter displays results for both the standard core-corona configuration and the corona-only (``no-core") scenario. In all cases, that are statistically significant, jet-like events (low spherocity) have narrower balance function widths compared to isotropic events (high spherocity), consistent with back-to-back partonic scatterings producing more collimated pairs. 

Among the particle species studied, pions exhibit the broadest balance function widths, because their large yields and substantial resonance decay contributions introduce additional kinematic separation between balancing partners. Protons show intermediate widths, shaped by baryon number conservation and their different hadronization dynamics. Kaon shows the narrowest widths, as strangeness conservation and resonce deacy $\phi \rightarrow K^{+}K^{-}$  tend to keep balancing charges more localized in phase space.
Significant and instructive differences, however, arise between the models: In \pythia, the balance function widths decrease monotonically with increasing event multiplicity, regardless of particle species or event shape. This trend suggests that, as event activity grows, balancing charges are created in closer proximity, indicating strong local charge conservation and reduced spatial diffusion in a scenario dominated by multi-parton interactions and independent string fragmentation. Moreover, the distinction between jet-like and isotropic events becomes less pronounced at high multiplicity, as the large number of overlapping soft and hard processes together with enhanced color reconnection effects tends to wash out the topological differences in the underlying correlation structure.

In contrast, the \epos model shows a weaker multiplicity dependence, especially for the core-corona scenario. For $\Delta y$ (upper panels in Figure~\ref{fg:eposwidthModel}), the balance function widths for the core-corona case are larger than those for the no-core (corona-only) scenario, especially at high multiplicity. The reduced width in $\Delta y$ observed in the corona-only (no-core) case arises from the dominance of independent string fragmentation, which lacks the collective, medium-driven longitudinal expansion and diffusion present in the core. 
 Without hydrodynamic effects, longitudinal expansion and charge transport are suppressed, resulting in narrower correlations in rapidity. In contrast, the presence of a hydrodynamic core introduces enhanced longitudinal flow and diffusion, increasing the separation between balancing charges and thereby broadening the balance function in $\Delta y$. In contrast, the situation in $\Delta\phi$ is reversed: the ``no-core" scenario consistently yields a broader balance function in azimuth compared to the core-corona case. The absence of strong radial flow in the corona-only events means that balancing charges are distributed more diffusely in azimuthal angle. However, the presence of a hydrodynamic core introduces strong collective radial expansion, which collimates the emission of particles, thus narrowing the balance function in $\Delta\phi$. In $\Delta \phi$ (lower panels), the core-corona widths are systematically lower than those for the corona-only scenario.  Across both \pythia and \epos models, a clear dependence of the balance function widths on event spherocity is observed: jet-like (low spherocity) events generally yield narrower widths than isotropic (high spherocity) events, especially for pions and kaons. In \pythia, this spherocity dependence is most pronounced at low multiplicity but diminishes with increasing multiplicity. For \epos, the distinction between jet-like and isotropic events persists at high multiplicity in the core-corona scenario, again particularly for pions and kaons, suggesting that the presence of a hydrodynamic core maintains sensitivity to event shape. However, for protons, at the highest multiplicities studied, the balance function widths converge and show minimal dependence on event shape in all three classes. This behavior likely reflects the stronger constraints from baryon number conservation and the different hadronization dynamics for protons compared to lighter hadrons. In summary, the combined study of balance function widths in \pythia and \epos, with and without hydrodynamic core, demonstrates the sensitivity of these observables to both microscopic (fragmentation) and macroscopic (collective flow) mechanisms in high-multiplicity pp collisions. The interplay between event multiplicity, event shape, and particle species reveals distinct features that distinguish models with collective effects from those relying on independent partonic interactions.

\section{Summary}
\label{summary}
In this study, we have systematically explored the charge balance functions for pions, kaons, and protons in proton-proton (pp) collisions at $\sqrt{s} = 13$~TeV, focusing on the interplay between event activity (charged-particle multiplicity) and event topology (spherocity) using the \pythia and \epos event generators. This reveals pronounced model-dependent trends in the widths of the balance functions as functions of relative rapidity ($\Delta y$) and relative azimuthal angle ($\Delta\phi$), within $0.2 < p_{\mathrm{T}} < 2.0$~GeV and $|y| \leq 2.4$. The \pythia simulations exhibit a clear monotonic narrowing of the balance function widths with increasing multiplicity for all particle species and angular variables. This behavior is indicative of stronger local charge conservation and reduced spatial diffusion as event activity increases, characteristic of a particle production scenario dominated by independent string fragmentation and multi-parton interactions. The observed ordering of widths, broadest for pions, followed by protons and then kaons, reflects the combined influence of particle mass, conservation laws, and hadronization dynamics. Jet-like events consistently yield narrower widths compared to isotropic events, emphasizing the collimating effect of back-to-back partonic scatterings in low-spherocity topologies. This indicates that the topological structure of the event, reflecting the relative contributions of hard scatterings and soft processes, has a direct and significant impact on the spatial and angular separation of balancing charges.

Notably, for kaons and protons in \epos, the inclusion of a hydrodynamic core leads to a distinct pattern in the balance function widths. In the $\Delta y$ direction, the core-corona scenario yields systematically broader widths compared to the no-core case, indicating enhanced longitudinal diffusion and collective flow effects. For $\Delta\phi$, the core-corona widths are significantly narrower than those in the no-core scenario, reflecting the strong collimation of balancing charges due to collective radial flow. This reversal between $\Delta y$ and $\Delta\phi$ widths is particularly pronounced for protons, suggesting a stronger coupling of baryon number transport to collective expansion. However, this trend is not observed for pions, where the differences between the core-corona and no-core scenarios are less marked. These results highlight the balance function’s unique sensitivity to both microscopic and collective effects in small collision systems. Our findings underscore the need for future experimental measurements that combine particle identification with event shape selection to further probe the emergence of collectivity in high-multiplicity pp at the LHC.

\begin{acknowledgements}
SCB and AK acknowledge the support under the INFN postdoctoral fellowship.
\end{acknowledgements}


\bibliographystyle{utphys}
\bibliography{reference}

\end{document}